\documentclass[aps,showpacs,twocolumn,twoside,groupedaddress]{revtex4}
\usepackage{graphicx}
\usepackage{epsfig}
\usepackage{epsf}
\usepackage{amssymb}
\usepackage{amsmath}
\usepackage[toc]{appendix}

\begin{document}

\title{Dynamical theory of single photon transport in a one-dimensional waveguide coupled to identical and non-identical emitters}

\author{Zeyang Liao$^{1}$\footnote{zeyangliao@physics.tamu.edu},  Hyunchul Nha$^{2}$, and M. Suhail Zubairy$^{1}$}

\affiliation{$^1$Institute for Quantum Science and Engineering (IQSE) and Department of Physics and Astronomy, Texas A$\&$M University, College Station, TX 77843-4242, USA \\
$^2$Department of Physics, Texas A$\&$M University at Qatar, Education City, P.O. Box 23874, Doha, Qatar 
}

\begin{abstract}
We develop a general dynamical theory for studying a single photon transport in a one-dimensional (1D) waveguide coupled to multiple emitters which can be either identical or non-identical. In this theory, both the effects of the waveguide and non-waveguide vacuum modes are included. This theory enables us to investigate the propagation of an emitter excitation or an arbitrary single photon pulse along an array of emitters coupled to a 1D waveguide. The dipole-dipole interaction induced by the non-waveguide modes, which is usually neglected in the literatures, can significantly modify the dynamics of the emitter system as well as the characteristics of output field if the emitter separation is much smaller than the resonance wavelength.  Non-identical emitters can also strongly couple to each other if their energy difference is smaller than or of the order of the dipole-dipole energy shift. Interestingly, if their energy difference is close but non-zero, a very narrow transparency window around the resonance frequency can appear which does not occur for identical emitters. This phenomenon may find important applications in quantum waveguide devices such as optical switch and ultra narrow single photon frequency comb generator. 
\end{abstract}

  \maketitle

\section{Introduction}

Photonic structure with reduced dimensions, such as 1D  photonic waveguide, can not only enhance the photon-emitter interaction but also guide the photons, which may find important applications in quantum devices and quantum information \cite{Noda2007, Leistikow2011}. A number of systems can be treated as a 1D waveguide such as optical nanofibers \cite{Dayan1062}, photonic crystal with line defects \cite{Englund2007}, surface plasmon nanowire \cite{Akimov402}, and superconducting microwave transmission lines \cite{ Wallraff2004, Abdumalikov193601, Hoi073601, Hoi263601, Loo1494}. These 1D systems are also excellent platforms for studying many-body physics since the interaction between the emitters induced by the waveguide modes can be long-range \cite{Douglas2015}. Strong photon-photon interaction may be also achieved in these systems \cite{Shen153003, Zheng2011, Roy2011a, Shi063803, Fang053845}. In analogy with ``cavity quantum eletrodynamics (QED)'', this system is usually termed as ``waveguide QED'' \cite{Liao063004}.

The stationary results of the photon transport in a waveguide-QED system, including a single photon or multiple photons interacting with a single emitter or multiple emitters, have been extensively studied based on the Bethe-ansatz approach \cite{Shen2005a, Shen2005b, Shen2007, Yudson2008, Tsoi2008, Zheng063816}, Lippmann-Schwinger scattering theory \cite{Huang2013, Li063810}, input-output theory \cite{Fan063821, Lalumiere2013, Xu043845},  Lehmann-Symanzik-Zimmermann reduction approach \cite{Shi205111}, and the diagrammatic method \cite{Pletyukhov095028}. In addition, dynamical theories, which allow to study the real time evolution of the emitter excitations and photon pulse, have also been studied \cite{Rephaeli2010, Chen2011, Liao2015, Baragiola2012, Shi053834}. Many applications of the waveguide-QED system have been proposed such as highly reflecting mirrors \cite{Zhou2008, Chang2012}, single-photon diode \cite{Menon2004, Shen173902}, efficient single-photon frequency converter \cite{Bradford103902, Bradford043814, Yan2013, Sun2014}, single-photon transistor \cite{Chang2007, Witthaut2010, Tiecke2014, Javadi2015}, photonic quantum gate \cite{Ciccarello2012, Zheng2013b, Paulisch2016}, and single photon frequency comb generator \cite{Liao2016}.

In the previous calculations \cite{Liao063004, Shen2005a, Shen2005b, Shen2007, Yudson2008, Tsoi2008, Zheng063816, Huang2013, Li063810, Fan063821, Lalumiere2013, Xu043845, Shi205111, Pletyukhov095028, Rephaeli2010, Chen2011, Liao2015, Baragiola2012, Shi053834}, the effect of the non-waveguide vacuum modes is included by simply adding a phenomenological decay factor in the Hamiltonian. This approximation is valid when the emitter separation is of the order of or larger than the resonant wavelength. However, it was recently shown that cold atoms can be trapped around a 1D waveguide even in the subwavelength region \cite{Vetsch2010, Hung083026, Tudela2015}. Quantum dots array with subwavelength separation can be also engineered \cite{Wang2006}. If the emitter separation is much smaller than the resonance wavelength, the emitter dipole-dipole interaction induced by the non-waveguide vacuum modes cannot be neglected \cite{Dicke1954, Ficek2004, Liao2012, Liao2014}.  In addition,  the emitters may have different transition frequencies due to the inhomogeneous local fields or nonuniform impurities \cite{Kim2011}, which is seldom considered in this system.

In this paper, going beyond earlier works, we develop a dynamical theory for single photon transport in a 1D waveguide-QED system where the emitters can be either identical or nonidentical and both the effects of the waveguide and the non-waveguide vacuum modes are included. When the emitter separation is much smaller than the resonant wavelength, the emitter dynamics and emission spectra can be significantly modified by the dipole-dipole interaction induced by the non-waveguide vacuum modes. From the modifications of the reflection and transmission spectra, we can clearly compare the results with and without the dipole-dipole interaction induced by the non-waveguide vacuum modes. We find that the dipole-dipole interaction induced by the non-waveguide vacuum modes can induce photon transparency in the waveguide system. In  addition, we also show that emitters with different transition frequencies can also significantly couple to each other and induce remarkable coherence effects. From the emission spectra we can quantify the effects of the dipole-dipole interaction between emitters with different transition frequencies, and we show the transition from coupled emitters to independent emitters as their energy difference increases. Interestingly, when the energy difference between the emitters is close but nonzero, a very narrow transparency window can occur around the resonance frequency. Similar effect has been studied in an ensemble of atoms based on semiclassical and mean-field theory and it is named as ``dipole-dipole induced eletromagnetic transparency (DIET)'' \cite{Joseph163603}. Here, we provide an ab initio calculations for the DIET and this phenomenon may be easier to experimentally observe in our system. Our theory here may provide an important tool for studying many-body physics and designing new waveguide-based quantum devices.

This paper is organized as follows. In Sec. II, we derive dynamical equations for a single photon transport in a 1D waveguide coupled to identical or non-identical emitters including the effects of the non-waveguide photon modes. We also derive the reflection and transmission photon spectra of this system. In Sec. III, we compare the results with and without including the dipole-dipole interaction induced by the non-waveguide photon modes in the cases that one emitter is initially excited or one single photon pulse is incident. By calculating the emission spectrum difference, we quantify the effects of the dipole-dipole interaction induced by the non-waveguide photon modes.  In Sec. IV, we study the photon transport in the case of non-identical emitters where we show that DIET can occur in this system. We also show the transition from coupled emitters to independent emitters by increasing the emitter energy difference. In Sec. V, we study the results beyond the two-emitter system where we show that very narrow single photon frequency combs can be generated. Finally, we summarize our results.

\section{Model and theory}

\begin{figure}
\includegraphics[width=0.9\columnwidth]{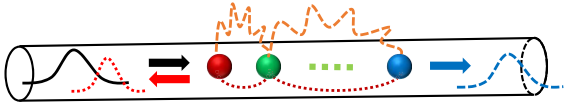}
\caption{(Color online)  Single photon transport in a 1D waveguide coupled to multiple identical or non-identical emitters. The emitters can couple to each other via the waveguide and non-waveguide photon modes. Black solid curve: incident field; Red dotted curve: reflected field; Blue dashed curve: transmitted field.}
\end{figure}

\subsection{Emitter excitation dynamics}

We consider a single photon transport in a 1D waveguide coupled to multiple quantum emitters which may have different transition frequencies (Fig. 1). The emitters can interact with the waveguide and non-waveguide photon modes. The interaction Hamiltonian in the rotating wave approximation is given by \cite{Scully2001}
\begin{multline} \label{eq1}
H=\hbar\sum_{j=1}^{N_a}\sum_{k}(g_{k}^{j}e^{ikz_{j}}\sigma_{j}^{+}a_{k}e^{-i\delta\omega_{k}^{j}t}+H.c.) \\ +\hbar\sum_{j=1}^{N_a}\sum_{\vec{q}_{\lambda} }(g_{\vec{q}_{\lambda} }^{j}e^{i\vec{q}\cdot \vec{r}_{j}}a_{\vec{q}_{\lambda}}\sigma_{j}^{+}e^{-i\delta\omega_{\vec{q}_{\lambda}}^{j}t}+H.c.) 
\end{multline}
where the first term is the coupling between the quantum emitters and the waveguide photon modes, and the second term is the coupling between the quantum emitters and the non-waveguide photon modes. Here, $N_a$ is the number of quantum emitters, $\delta\omega_{k}^{j}=\omega_{k}-\omega_{j}$ ($\delta\omega_{\vec{q}_{\lambda}}=\omega_{\vec{q}_\lambda}-\omega_{j}$) is the detuning between the transition frequency of the $j$th emitter $\omega_{j}$ and the frequency  $\omega_{k}$ ($\omega_{\vec{q}_\lambda}$) of the guided photon with wavevector $k$ (non-waveguide photon modes with polarization $\lambda$ and wavevector $\vec{q}$). If $\omega_{j}$ is far away from the cutoff frequency of the photonic waveguide and the waveguide photon has a narrow bandwith, we can linearize the waveguide photon dispersion relation as  $\delta\omega_{k}^{j}=(|k|-k_{j})v_{g}$ where $k_{j}$ is the wave vector at frequency $\omega_{j}$ and $v_{g}$ is the group velocity \cite{Shen023837}. $\sigma_{j}^{+}=|e\rangle_j\langle g|  (\sigma_{j}^{-}=|g\rangle_j\langle e|)$ is the raising (lowering) operator of the $j$th emitter with position $r_{j}$ ($z_{j}$ is its $z$ component along the waveguide direction). $a_{k}^{\dagger } (a_{k}^{-})$ and $a_{\lambda, \vec{q}}^{\dagger } (a_{\lambda, \vec{q}}^{-})$ are the creation (annihilation) operators of a guided photon and a non-guided photon. $g_{k}^{j}=\vec{\mu}_j\cdot\vec{E}_{k}(\vec{r}_j)/\hbar$ is the coupling strength between the $j$th emitters and the guided photon modes with $\vec{\mu}_j$ being the transition dipole moment of the $j$th emitter and $\hbar$ being the Planck constant, and $g_{\lambda ,\vec{q} }^{j}=\vec{\mu}_j\cdot\vec{E}_{\lambda ,\vec{q}}(\vec{r}_j)/\hbar $ is the coupling strength between the $j$th quantum emitter and the non-guided photon modes.

For a single photon excitation, the quantum state of the system at an arbitrary time can be expressed as 
\begin{multline} \label{eq2}
|\Psi(t)\rangle = \sum_{j=1}^{N_a}\alpha_{j}(t)|e_{j},0_{k}, 0_{\vec{q}_\lambda}\rangle+\sum_{k}\beta_{k}(t)|g,1_{k},0_{\vec{q}_{\lambda}}\rangle \\
+ \sum_{\vec{q}_{\lambda}}\gamma_{\vec{q}_{\lambda}}(t)|g,0_{k},1_{\vec{q}_{\lambda}}\rangle
\end{multline}
where $|e_{j},0_{k}, 0_{\lambda ,\vec{q}}\rangle$ is the state in which only the $j$th emitter is excited with zero waveguide and non-waveguide photons, $|g,1_{k},0_{\lambda ,\vec{q}}\rangle$ is the state in which all the emitters are in the ground state and one waveguide photon is generated with zero non-waveguide photon, and $|g,0_{k},1_{\vec{q}_{\lambda}}\rangle$ is the state where all the emitters are in the ground state and one non-waveguide photon is generated with zero photon being in the waveguide. $\alpha_{j}(t), \beta_{k}(t),$ and $\gamma_{\vec{q}_\lambda}(t)$ are the corresponding amplitudes at time $t$. 

From the Schr\"{o}dinger equation $i\hbar\partial _t|\Psi(t)\rangle=H|\Psi(t)\rangle$ with Hamiltonian given by Eq. (1) and the quantum state given by Eq. (2), we obtain the following dynamical equations for the probability amplitudes
\begin{align}
i\dot{\alpha_{j}}(t)=&\sum_{k}g_{k}^{j}e^{ikz_{j}-i\delta\omega_{k}^{j}t}\beta_{k}(t) +\sum_{\vec{q}_{\lambda}}g_{\vec{q}_{\lambda}}^{j} e^{i\vec{q}\cdot\vec{r}_{j}-i\delta\omega_{\vec{q}_{\lambda}}^{j}t}\gamma_{\vec{q}_{\lambda}}(t),  \\
i\dot{\beta}_{k}(t)=&\sum_{j=1}^{N_a}g_{k}^{j*}e^{-ikz_{j}}e^{i\delta\omega_{k}^{j}t}\alpha_{j}(t), \\
i\dot{\gamma}_{\vec{q}_{\lambda}}(t)=&\sum_{j=1}^{N_a}g_{\vec{q}_{\lambda}}^{j*}e^{-i\vec{q}\cdot\vec{r_{j}}}e^{i\delta\omega_{\vec{q}_{\lambda}}^{j}t}\alpha_{j}(t).
\end{align}
Integrating Eqs. (4) and (5), we obtain the formal solutions of the photon amplitudes which are given by
\begin{align}
\beta_{k}(t)&=\beta_{k}(0)-i\sum_{j=1}^{N_a}g_{k}^{j*}e^{-ikz_{j}}\int_{0}^{t}\alpha_{j}(t')e^{i\delta\omega_{k}^{j}t'}dt', \\
\gamma_{\vec{q}_\lambda}(t)&=\gamma_{\vec{q}_\lambda}(0)-i\sum_{j=1}^{N_a}g_{\vec{q}_\lambda}^{j*}e^{-i\vec{q}\cdot \vec{r}_{j}}\int_{0}^{t}\alpha_{j}(t')e^{i\delta\omega_{\vec{q}_{\lambda}}^{j}t'}dt',  
\end{align}
where $\beta_{k}(0)$ is the initial guided photon amplitude, and $\gamma_{\vec{q}_\lambda}(0)$ is the initial non-guided photon amplitude. In this paper, we assume that there is no photon in the non-waveguide photon modes initially, i.e., $\gamma_{\vec{q}_\lambda}(0)=0$. Inserting Eqs. (6) and (7) into Eq. (3), we obtain
\begin{multline}
\dot{\alpha_{j}}(t)=-i\sum_{k}g_{k}^{j}e^{ikz_{j}}e^{-i\delta\omega_{k}^{j}t}\beta_{k}(0) \\
-\sum_{l=1}^{N_a}\sum_{k}g_{k}^{j}g_{k}^{l*}e^{ik(z_{j}-z_{l})}\int_{0}^{t}dt' \alpha_{l}(t')e^{i\delta\omega_{k}^{l}t'}e^{-i\delta \omega_{k}^{j}t} \\
-\sum_{l=1}^{N_a}\sum_{\vec{q},\lambda}g_{\vec{q}_\lambda}^{j}g_{\vec{q}_\lambda}^{l*}e^{i\vec{q}\cdot(\vec{r}_{j}-\vec{r}_{l})}\int_{0}^{t}dt' \alpha_{l}(t')e^{i\delta\omega_{\vec{q}_{\lambda}}^{l}t'} e^{-i\delta \omega_{\vec{q}_{\lambda}}^{j}t},
\end{multline}
where the first term is the excitation by the incident waveguide photon, the second and the third terms are the coupling between the emitters induced by the waveguide vacuum modes and non-waveguide vacuum modes, respectively. 

By summing over the second and third terms of Eq. (8) using the Weisskopf-Wigner approximation, we can obtain closed dynamical evolution equations of the emitters given by (see Appendix)
\begin{multline}
\dot{\alpha}_{j}(t)=b_{j}(t)-\sum_{l=1}^{N_a} [V_{jl}^{(w)}e^{ik_{l} z_{jl}} \alpha_{l}(t-\frac{z_{jl}}{v_{g}})\\+V_{jl}^{(nw)}e^{ik_{l} r_{jl}} \alpha_{l}(t-\frac{r_{jl}}{v_{g}})]e^{i\Delta \omega_{jl} t}, 
\end{multline}
with $j=1,\cdots,N_{a}$. In Eq. (9),
\begin{equation}
b_{j}(t)=\frac{-i}{2\pi}\sqrt{\frac{\Gamma_j v_{g}L}{2}}e^{ik_{a}z_{j}}e^{i\Delta k_{j}v_{g}t}\int_{-\infty }^{\infty }\beta_{0}(\delta k)e^{i\delta k(r_{j}-v_{g}t)}d\delta k 
\end{equation} 
is the incident photon excitation,  
$V_{jj}^{(w)}=\Gamma_{j}/2$ with $\Gamma_{j}=2L|g_{k_{a}}^{j}|^2/v_{g}$ the decay rate of the $j$th emitter due to the waveguide vacuum modes ($L$ is the quantization length of the waveguide, ) and $V_{jj}^{(nw)}=\gamma_{j}/2$ with $\gamma_j=k_j^{3}\mu_{j}^{2}/3\pi\hbar\epsilon _{0}V$ the spontaneous decay rate of the $j$th emitter due to the non-waveguide vacuum modes ($\epsilon _{0}$ is vacuum permittivity and $V$ is the quantization volume). For $j\neq l $, $V_{jl}^{(w)}=\sqrt{\Gamma_{j}\Gamma_{l}}/2$ is the dipole-dipole coupling strength due to the waveguide vacuum modes with $z_{jl}=|z_{j}-z_{l}|$ and
\begin{equation}
V_{jl}^{(nw)}=\frac{3\sqrt{\gamma_{j}\gamma_{l}}}{4}[\frac{-i}{k_{a}r_{jl}}+\frac{1}{(k_{a}r_{jl})^2}+\frac{i}{(k_{a}r_{jl})^3}]
\end{equation}
is the dipole-dipole interaction due to the non-waveguide photon modes with $r_{jl}=|\vec{r}_{j}-\vec{r}_{l}|$ where we assume the emitter dipole moment is perpendicular to the emitter chain. $\Delta k_{j}=k_{j}-k_{a}$ where $k_{a}$ can be chosen as the average emitter wavevector. The term $e^{i\Delta \omega_{a}^{jl} t}$ is due to the energy difference between the $j$th and $l$th emitters with $\Delta \omega_{jl}=(k_{j}-k_{l})v_{g}$. If the emitters have the same transition frequencies, the equation describes the case for identical emitters. Using Eq. (9) and the given initial conditions, we can calculate the real-time evolution of the emitter system with arbitrary configurations.  

\subsection{Emission spectra}

In addition to the emitter excitation, we can also calculate the emission photon spectrum. The waveguide photon spectrum at arbitrary time can be also calculated by Eq. (6) after obtaining the emitter excitation $\alpha_{j}(t)$. For simplicity, we assume $\Gamma_{j}=\Gamma$ and $\gamma_{j}=\gamma$ in the following calculations. Particularly, long time after the interaction, i.e., $t\rightarrow \infty$, the reflection and transmission waveguide photon spectra are given by
\begin{align}
\beta_{R}(\delta k)&=-i\sqrt{\frac{\Gamma v_{g}}{2L}}\sum_{j=1}^{N_{a}}e^{ikz_{j}}\int_{0}^{\infty}\alpha_{j}(t')e^{i(\delta k-\Delta k_{j}) v_{g}t'}dt', \\
\beta_{T}(\delta k)&=\beta_{0}(\delta k) \nonumber \\ &-i\sqrt{\frac{\Gamma v_{g}}{2L}}\sum_{j=1}^{N_{a}}e^{-ikz_{j}}\int_{0}^{\infty}\alpha_{j}(t')e^{i(\delta k-\Delta k_{j}) v_{g}t'}dt',
\end{align}
where $k=k_{a}+\delta k$.

We can perform the Fourier transformation and define 
\begin{equation}
\chi _{j}(\delta k)=\int_{-\infty}^{\infty}\alpha_{j}(t)\Theta(t)e^{i\delta kv_{g}t}dt
\end{equation}
where $\Theta(t)$ is the unit step function with $\Theta(t)=1$ for $t\geq 0$ and $\Theta(t)=0$ for $t<0$. The photon spectra shown in Eqs. (12) and (13) can then be rewritten as
\begin{align}
\beta_{R}(\delta k)&=-i\sqrt{\frac{\Gamma v_{g}}{2L}}\sum_{j=1}^{N_{a}}e^{ikz_{j}} \chi _{j}(\delta k-\Delta k_{j}),  \\
\beta_{T}(\delta k)&=\beta_{0}(\delta k)-i\sqrt{\frac{\Gamma v_{g}}{2L}}\sum_{j=1}^{N_{a}}e^{-ikz_{j}}\chi _{j}(\delta k-\Delta k_{j}).
\end{align}

Therefore, to calculate the photon spectrum we need to first calculate $\chi _{j}(\delta k)$. For this purpose, we perform the inverse Fourier transformation 
\begin{equation}
\alpha_{j}(t)\Theta(t)=\frac{v_{g}}{2\pi}\int_{-\infty}^{\infty}\chi _{j}(\delta k)e^{-i\delta kv_{g}t}d\delta k.
\end{equation}
Next, using the relation 
$d/dt[\alpha_{j}(t)\Theta(t)]=\dot{\alpha}_{j}(t)\Theta(t)+\alpha_{j}(0)\delta (t)$, we obtain a set of linear equations for $\chi _{j}(\delta k)$ from Eq. (9) which are given by
\begin{multline}
-i(\delta k-\Delta k_{j})v_{g}\chi _{j}(\delta k-\Delta k_{j})\\=A_{j}(\delta k-\Delta k_{j})  -\sum_{l=1}^{N_a}V_{jl}e^{ikz_{jl}}\chi_{l}(\delta k-\Delta k_{l}).
\end{multline}
Here, for simplicity, we assumed that the emitters are all aligned with the waveguide and we have $r_{jl}=z_{jl}$ and $V_{jl}=V_{jl}^{(w)}+V_{jl}^{(nw)}$. In Eq. (18), $A_{j}(\delta k)=\alpha_{j}(0)+b_{j}(\delta k)$ where $\alpha_{j}(0)$ is the initial excitation of the $j$th emitter, and
\begin{equation}
b_{j}(\delta k)=-i\sqrt{\frac{\Gamma L}{2v_{g}}}\beta_{0}(\delta k+\Delta k_{j})e^{i(k_{j}+\delta k)z_{j}}
\end{equation}
is the initial waveguide photon spectrum.  

The solution of Eq. (18) can be calculated as
\begin{equation}
\chi _{j}(\delta k-\Delta k_{j})=\sum_{l=1}^{N_a}[M(\delta k)]_{jl}^{-1}A_{l}(\delta k-\Delta k_{l})
\end{equation}
where $M(\delta k)$ is an $N_{a}\times N_{a}$ matrix with matrix element given by 
\begin{equation}
[M(\delta k)]_{pq}=V_{pq}e^{ikz_{pq}}-i(\delta k-\Delta k_{p})v_{g}\delta_{pq}.
\end{equation}
From Eqs. (15), (16), and (20), we can calculate the reflection and the transmission spectra. For the case with one emitter excitation but without incident photons, we have the photon spectra to the left (``-'') and to the right (``+'') given by 
\begin{equation}
\beta_{\pm}(\delta k)=-i\sqrt{\frac{\Gamma v_{g}}{2L}}\sum_{j,l=1}^{N_{a}}\alpha_{l}(0)[M(\delta k)]^{-1}_{jl}e^{\mp ikz_{j}}.
\end{equation}
For a single incident photon pulse without any initial emitter excitation, we have the reflection and transmission spectra given by
\begin{align}
\beta_{R}(\delta k)&=-\frac{\Gamma}{2}\beta_{0}(\delta k)\sum_{j,l=1}^{N_{a}}[M(\delta k)]^{-1}_{jl}e^{ik(z_{j}+z_{l})}, \\
\beta_{T}(\delta k)&=\beta_{0}(\delta k)\Big\{1-\frac{\Gamma}{2}\sum_{j,l=1}^{N_{a}}[M_{j}(\delta k)]^{-1}_{jl}e^{ik(z_{l}-z_{j})}\Big\}.
\end{align}

For a single emitter case, it is readily obtained from Eq. (20) that   
\begin{equation}
\chi_{1}(\delta k)=\frac{\alpha_{1}(0)+b_{1}(\delta k)}{V_{11}-i\delta k v_{g}}
\end{equation}
where $V_{11}=(\Gamma+\gamma)/2$. When the emitter is initially excited and there is no input photon, i.e., $\alpha_{1}(0)=1$ and $b_{1}(\delta k)=0$, the spontaneous emission spectrum has the usual Lorentzian line shape.  For a single photon input with the emitter being initially in the ground state, i.e., $\alpha_{l}(0)=0$ and $b_{1}(\delta k)\neq 0$, the emission spectrum is a Lorentzian function modulated by the input photon spectrum.

In the following sections, we first study the effects of dipole-dipole interaction induced by the non-waveguide vacuum modes with two-emitter example. Then we study the effects of non-identical emitters with two-emitter example. Finally we study the case beyond two-emitter system.

\begin{figure*}
\includegraphics[width=0.9\columnwidth, bb=0 100 575 575]{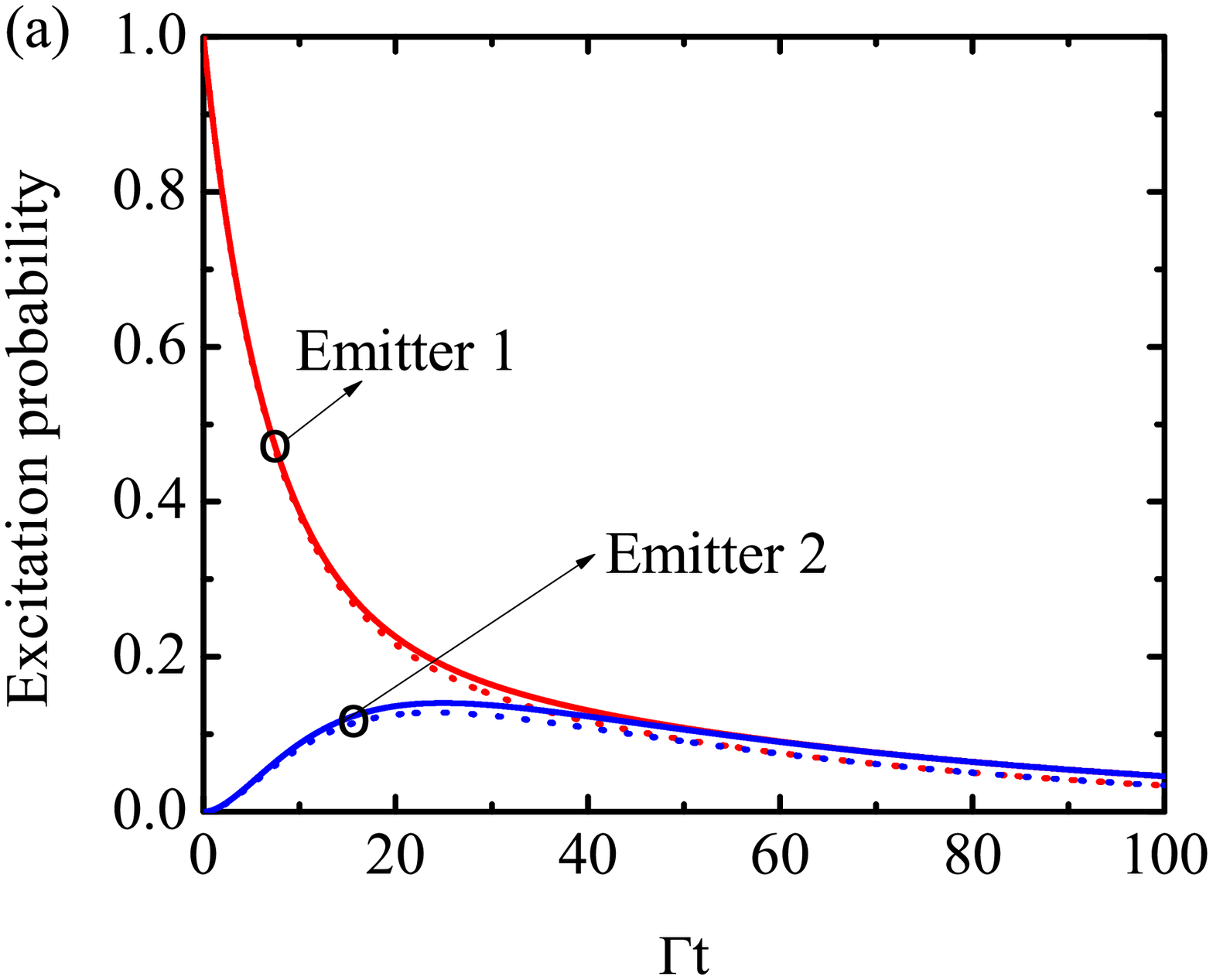}
\includegraphics[width=0.9\columnwidth, bb=0 100 575 575]{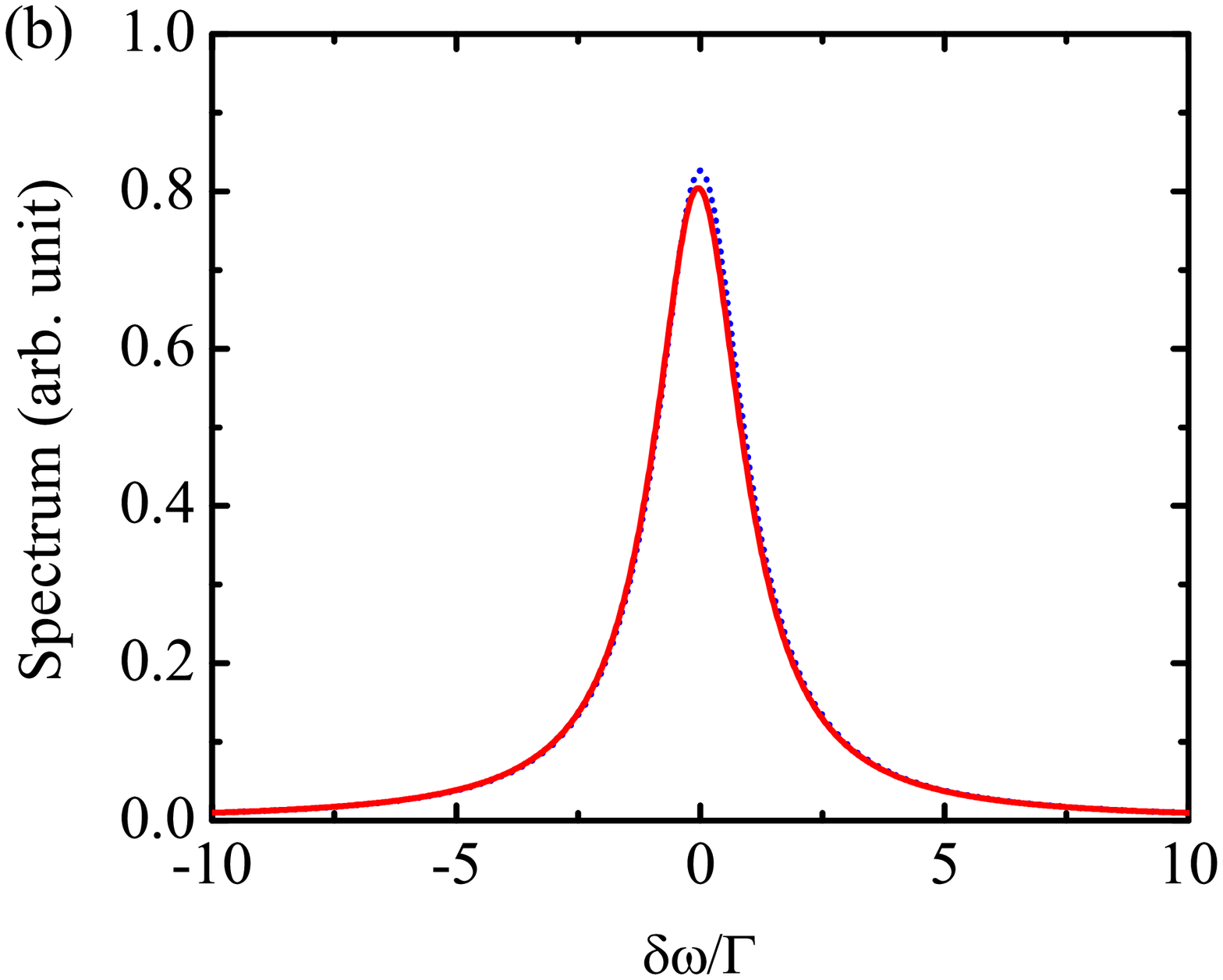}
\includegraphics[width=0.9\columnwidth, bb=0 100 575 575]{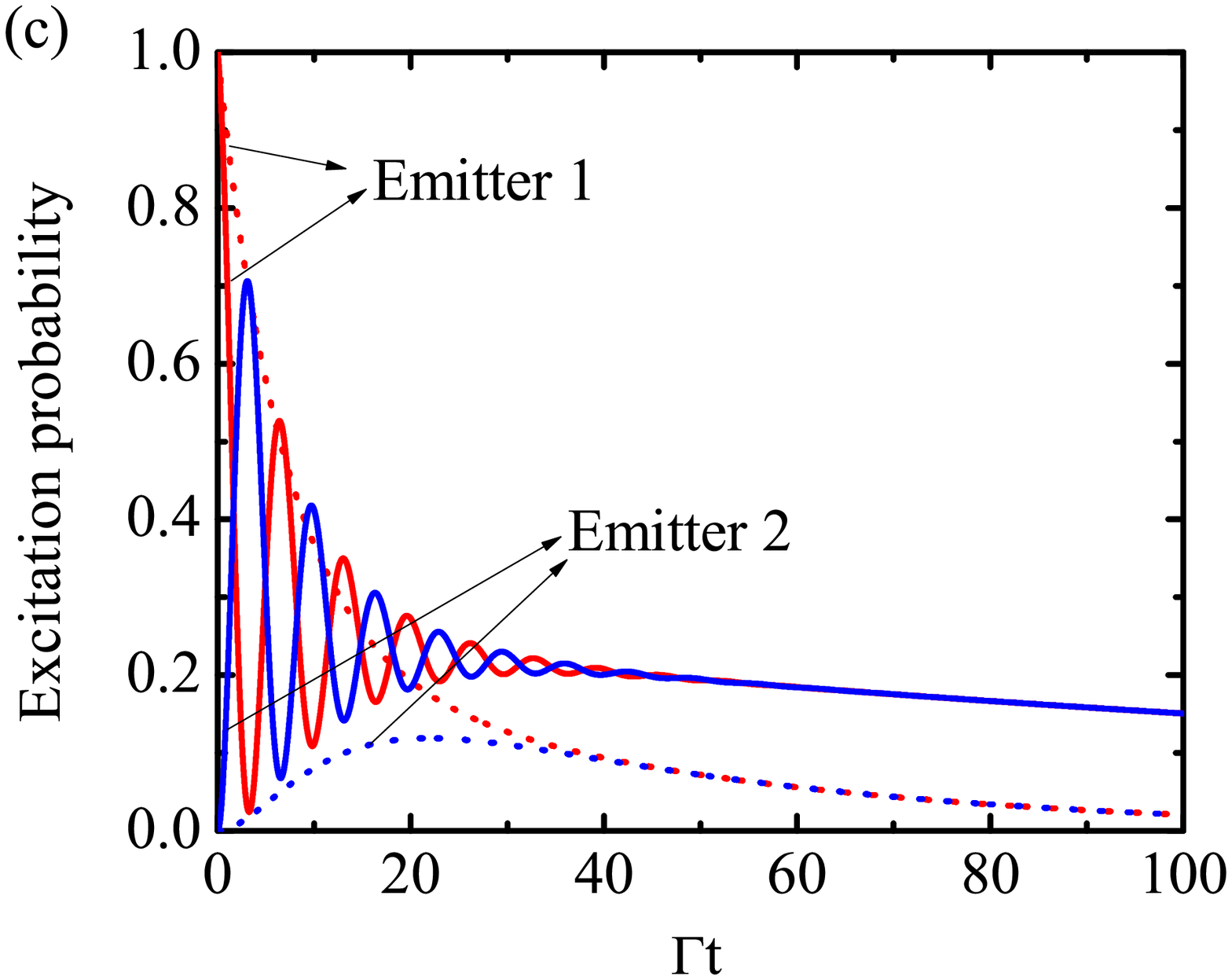}
\includegraphics[width=0.9\columnwidth, bb=0 100 575 575]{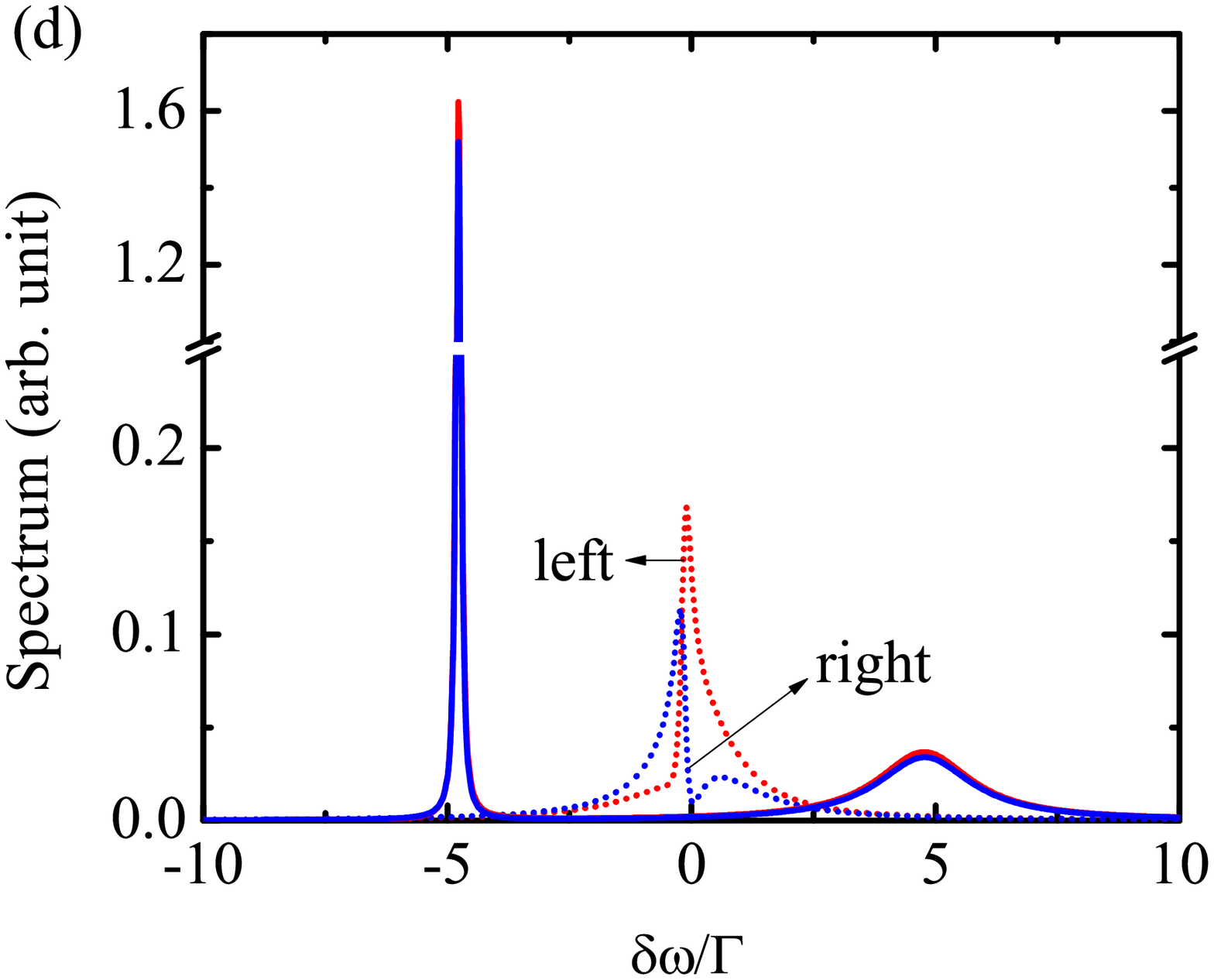}
\caption{(Color online) (a, c) Emitter excitation dynamics as a function of time.  (b, d) Emission photon spectra. Emitter 1 is initially excited and there is no incident photon.  $\alpha_{1}(0)=1, \alpha_{2}(0)=b_{1,2}(\delta k)=0, \gamma=0.2\Gamma$. (a, b) $a=0.5\lambda$. (c, d) $a=0.05\lambda$. Dotted lines are the results without $V_{12}^{(nw)}$, and the solid curves are the results with $V_{12}^{(nw)}$. }
\end{figure*}

\section{Effects of dipole-dipole interaction induced by the non-waveguide vacuum modes}

In this section, we consider the emitters to be identical and compare the results with and without including the dipole-dipole interaction induced by the non-waveguide vacuum mode. For identical emitters, we have $k_{j}=k_{a}$ for all emitters.

For a two-emitter system, the evolution of the emitters is given by
\begin{align}
\dot{\alpha}_{1}(t)&=b_{1}(t)-V_{11}\alpha_{1}(t)-V_{12}e^{ik_{a}z_{12}}\alpha_{2}(t-\frac{z_{12}}{v_{g}}), \\
\dot{\alpha}_{2}(t)&=b_{2}(t)-V_{22}\alpha_{2}(t)-V_{21}e^{ik_{a}z_{12}}\alpha_{2}(t-\frac{z_{12}}{v_{g}}),
\end{align}
where $V_{11}=V_{22}=(\Gamma+\gamma)/2$, $V_{12}=V_{21}=V_{12}^{(w)}+V_{12}^{(nw)}$, and $b_{1,2}(t)$ are given by Eq. (10). Due to the dipole-dipole coupling, the one excitation subspace is split into two eigenstates ($|+\rangle = (|eg\rangle +|ge\rangle)/\sqrt{2}$ and $|-\rangle = (|eg\rangle -|ge\rangle)/\sqrt{2}$) with the energy shifts given by $\pm \text{Im}[V_{12}e^{ik_{a}z_{12}}]$ and the decay rates given by $V_{11}\pm \text{Re}[V_{12}e^{ik_{a}r_{12}}]$. The $M(\delta k)$ matrix for calculating the emission spectra are given by
\begin{equation}
M(\delta k)=\begin{bmatrix} V_{11}-i\delta k v_{g} & V_{12}e^{ik_a z_{12}} \\ V_{21}e^{ik_a z_{12}} & V_{22}-i\delta k v_{g} \end{bmatrix}.
\end{equation}

\subsection{Emitter excitation propagation}

In this subsection, we consider the propagation of emitter excitation without incident photon pulse. We assume that the emitter on the left is initially in the excited state while the emitter on the right is initially in the ground state, $\alpha_{1}(0)=1, \alpha_{2}(0)=b_{1,2}(\delta k)=0$. In this case, the photon emission spectra to the left and to the right are given by 
\begin{align}
\beta_{-}(\delta k)&=-i\sqrt{\frac{\Gamma v_{g}}{2L}}e^{ikz_{1}}\frac{V_{11}-i\delta k v_{g}-V_{12}e^{2ikz_{12}}}{(V_{11}-i\delta k v_{g})^2-V_{12}^{2}e^{2ikz_{12}}}, \\
\beta_{+}(\delta k)&=-i\sqrt{\frac{\Gamma v_{g}}{2L}}e^{-ikz_{1}}\frac{V_{11}-i\delta k v_{g}-V_{12}}{(V_{11}-i\delta k v_{g})^2-V_{12}^{2}e^{2ikz_{12}}},
\end{align}
where $k=k_{a}+\delta k$.

We compare the emitter dynamics and the emission spectra in the cases with and without including the dipole-dipole interaction induced by the non-waveguide vacuum modes  ($V_{dd}^{(nw)}$). Here we study the cases of two emitter separations, i.e., $a=0.5\lambda$ and $a=0.05\lambda$.  The emitter excitations and the emission spectra when $a=0.5\lambda$ and $\gamma=0.2\Gamma$ are shown in Fig. 2(a) and 2(b), respectively. In this case, $V_{ii}=0.6\Gamma$, $V_{12}^{(w)}=0.5\Gamma$, and $V_{12}^{(nw)}=0.015\Gamma-0.043\Gamma i$. Without including $V_{12}^{(nw)}$, $\text{Im}(V_{12}e^{ik_{a}r_{12}})=0$ which gives zero energy shifts for the two eigenstates. The two decay rates are given by $1.1\Gamma$ and $0.1\Gamma$ corresponding to a superradiant and a subradiant state, respectively. With $V_{12}^{(nw)}$, the energy shifts are given by $\pm 0.043\Gamma$ and the decay rates are given by $1.115\Gamma$ and $0.085\Gamma$. The difference between the cases with and without $V_{12}^{(nw)}$ is very small. Indeed, from Fig. 2(a) and 2(b), we see that both the emitter excitations and the photon spectra are almost the same with (solid curves) and without (dotted curves) including $V_{12}^{(nw)}$. The spectra in two directions are the same and they have Lorentzian line shapes. Hence, when the emitter separation is relatively large, we can safely neglect the effect of $V_{12}^{(nw)}$ \cite{Liao2015}. 

However, when the emitter separation is very small compared with the resonant wavelength, the results are quite different. The emitter excitations and the photon spectra when $a=0.05\lambda$ with $\gamma=0.2\Gamma$ are shown in Fig. 2(c) and 2(d), respectively. In this case, $V_{ii}=0.6\Gamma$, $V_{12}^{(w)}=0.5\Gamma$, and $V_{12}^{(nw)}=1.52\Gamma+4.36\Gamma i$. Without including $V_{12}^{(nw)}$ the energy shifts are $\pm 0.15\Gamma$ and the decay rates are given by $1.08\Gamma$ and $0.12\Gamma$. With $V_{12}^{(nw)}$, the energy shifts are $\pm 4.77\Gamma$ and the decay rates are given by $1.17\Gamma$ and $0.03\Gamma$. There are large differences due to $V_{12}^{(nw)}$ which can also be seen from Fig. 2(c) and 2(d). Without including $V_{12}^{(nw)}$, the emitter excitation dynamics are similar to the case when $a=0.5\lambda$.  However, with the effect of $V_{12}^{(nw)}$ the two emitters exchange energy many times until they have the same excitation probability and then decay slowly to the ground state. Since the decay rate of the subradiant eigenstate with $V_{12}^{(nw)}$ ($0.03\Gamma$) is smaller than that without $V_{12}^{(nw)}$ ($0.12\Gamma$), the emitter excitations last much longer with $V_{12}^{(nw)}$ than those without $V_{12}^{(nw)}$.  The spectra are also quite different. Without including $V_{12}^{nw}$, the emission spectra are peaked close to the resonance frequency with Fano-like line shapes \cite{Fano1961}.  With $V_{12}^{nw}$, the emission spectra are far away from the resonance frequency. The spectra of the left-moving and the right-moving fields are almost the same with one superadiant peak and one subradiant peak. Therefore, $V_{12}^{(nw)}$ can be a crucial factor to determine the characteristics of the waveguide system if the emission to the non-waveguide modes ($\gamma$) is not too small and the emitter separation is much smaller than the resonance wavelength.

\begin{figure*}
\includegraphics[width=0.9\columnwidth, bb=0 100 575 575]{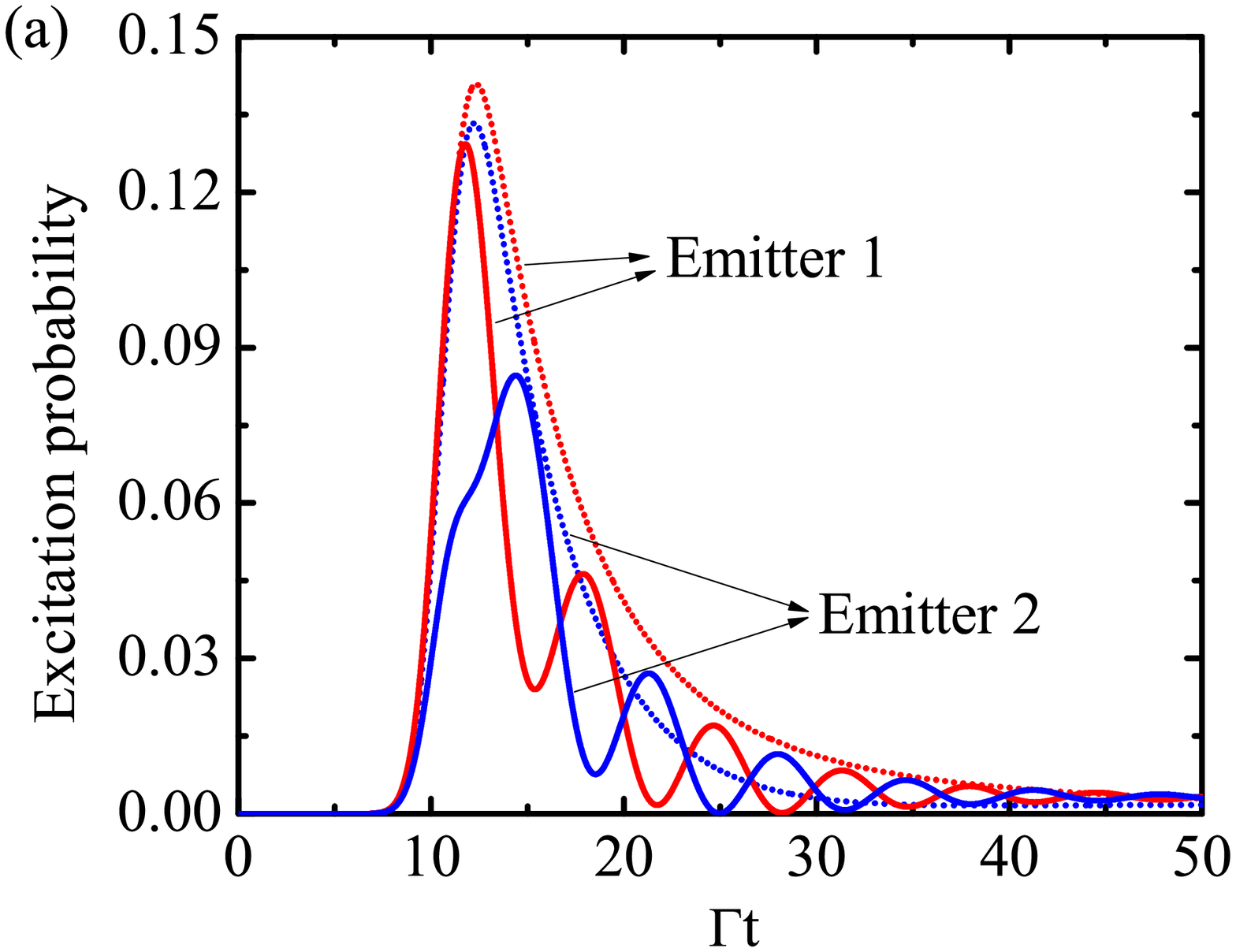}
\includegraphics[width=0.9\columnwidth, bb=0 100 575 575]{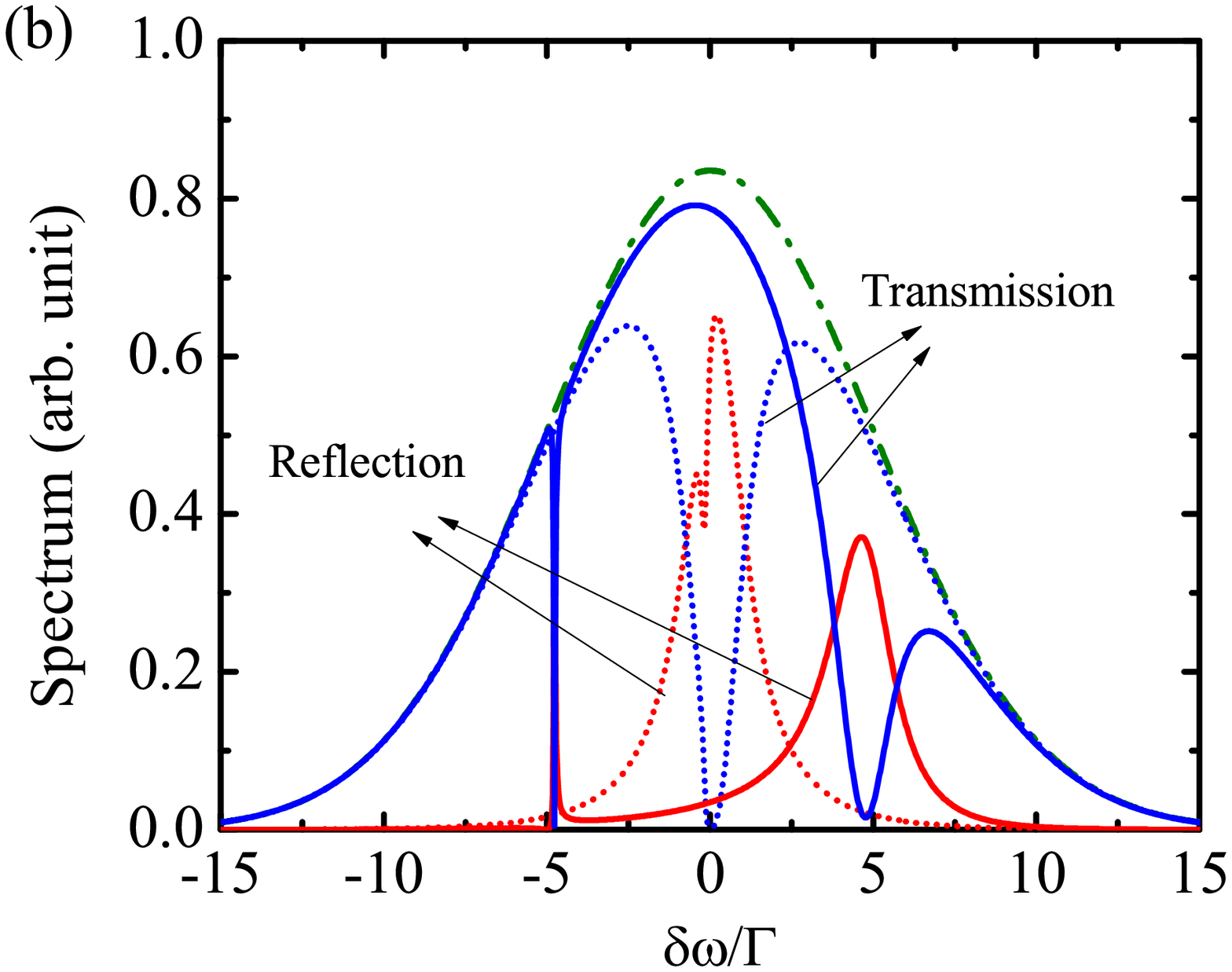}
\caption{(Color online) (a) Emitter excitation probabilities as a function of time excited by an incident photon pulse with (solid lines) and without (dotted lines) $V_{12}^{(nw)}$. (b) Emission photon spectra with and without $V_{12}^{(nw)}$. The green dahsed-dotted lines are the incident photon spectrum, the solid curves are the results with $V_{12}^{(nw)}$, and the dotted curves are the results without $V_{12}^{(nw)}$. Parameter: $\alpha_{1}(0)=\alpha_{2}(0)=0, \gamma=0.2\Gamma, \Delta_{0}v_{g}=10\Gamma$, $z_{1}=10/\Gamma$, and $a=0.05\lambda$. }
\end{figure*}

\subsection{Single photon transport}

Next, we consider the case when both emitters are initially in the ground state and there is a single incident photon pulse. The emitter dynamics are given by Eqs. (26) and (27). The reflection and transmission photon spectra are given by
\begin{align}
\beta_{R}(\delta k)&=-\frac{\Gamma}{2}e^{2ikz_{1}}\beta_{0}(\delta k)\nonumber \\ &\times\frac{(V_{11}-i\delta k v_{g})(1+e^{2ikz_{12}})-2V_{12}e^{2ikz_{12}}}{(V_{11}-i\delta k v_{g})^{2}-V_{12}^{2}e^{2ikz_{12}}}, \\
\beta_{T}(\delta k)&=\beta_{0}(\delta k)\Big [1-\frac{\Gamma}{2}\frac{2 (V_{11}-i\delta k v_{g})-V_{12}(1+e^{2ikz_{12}})}{(V_{11}-i\delta k v_{g})^{2}-V_{12}^{2}e^{2ikz_{12}}} \Big ],
\end{align}
where  $k=k_{a}+\delta k$. If the waveguide is so good that the non-waveguide modes are inhibited (i.e., $\gamma=0$), we have $V_{ii}=V_{ij}=\Gamma/2$ with $i,j=1,2$. In this case, it is not difficult to see that for resonance frequency we have $\beta_{R}(0)=-e^{2ikz_{1}}\beta_{0}(\delta k)$ and $\beta_{T}(0)=0$. Thus, without non-waveguide vacuum modes, the resonance frequency is completely reflected with a $\pi$ phase shift \cite{Liao2015}.

For illustration with the numerical examples, we assume that the photon pulse has a Gaussian shape with spectrum given by
\begin{equation}
\beta_{0}(\delta k)=\frac{(8\pi)^{1/4}}{\sqrt{\Delta_{0} L}}e^{-\frac{\delta k^2}{\Delta_{0}^2}},
\end{equation}
where $\Delta_{0}$ is the width in the $k$ space with the full width at half maximum of the spectrum being $\sqrt{2\ln 2}\Delta_{0} v_{g}$.  The single photon condition requires that $(L/2\pi)\int_{-\infty}^{\infty}|\beta_{0}(\delta k)|^{2}d\delta k=1$. With this Gaussian photon pulse, we have
\begin{equation}
b_{j}(t)=-i(8\pi)^{-1/4}\sqrt{\Gamma\Delta_{0}v_{g}}e^{ik_{0}z_{j}}e^{-\frac{1}{4}\Delta_{0}^{2}(z_{j}-v_{g}t)^2}
\end{equation}
in Eq. (9).
The emitter excitation as a function of time when $a=0.05\lambda $, $\gamma=0.2\Gamma$ is shown in Fig. 3(a) where we assume that $\Delta_{0}=10\Gamma$. The two dotted curves are the two-emitter excitations without including $V_{12}^{(nw)}$ and the two solid curves are those with $V_{12}^{(nw)}$. We can see that without $V_{12}^{(nw)}$ the two emitters are first exited and then deexcited with almost the same excitation dynamics. However, with  $V_{12}^{(nw)}$ the excitations of the two emitters can oscillate coherently after being excited by the photon pulse due to the strong dipole-dipole interaction between the two emitters. The emission spectra with and without $V_{12}^{(nw)}$ are also quite different as shown in Fig. 3(b). In the figure, the two dotted curves are the reflection and transmission spectra without $V_{12}^{(nw)}$ while the solid curves are those with $V_{12}^{(nw)}$.  Without $V_{12}^{(nw)}$, the resonant frequency is significantly reflected with negligible transmission. However, with $V_{12}^{(nw)}$, the resonant frequency can almost transmit with two reflection peaks far away from the resonant frequency. This is the phenomenon of dipole-dipole induced electromagnetic transparency (DIET).  Compared to the usual eletromagnetic induced transparency (EIT) where the transparency is caused by a strong pumping field \cite{Harris1990}, here the transparency is induced by the strong dipole-dipole interaction between the emitters. The strong dipole-dipole interaction can significantly shift the eigenenergy of the system and therefore the resonance frequency can be transmitted. This phenomenon may be used as optical switch \cite{Chang2007, Witthaut2010, Tiecke2014}. By controlling the emitter separation, we can control the dipole-dipole interaction between the emitters to control the transmission of the photons. However, in practice it is not easy to tune the emitter separation. In Sec. IV (A), we show that DIET can be achieved by simply tuning the emitter transition frequency which should be more convenient. The reflection occurs at the frequencies far away from the resonant frequency with one peak being superradiant peak while the other being subradiant peak  similar to Fig. 2(d). This example again shows that for small emitter separation the dipole-dipole interaction induced by the non-waveguide vacuum modes can play a non-trivial role if $\gamma$ is not too small compared with $\Gamma$.

\subsection{Spectrum difference with and without $V_{dd}^{(nw)}$}

\begin{figure}
\includegraphics[width=0.9\columnwidth, bb=0 100 575 575]{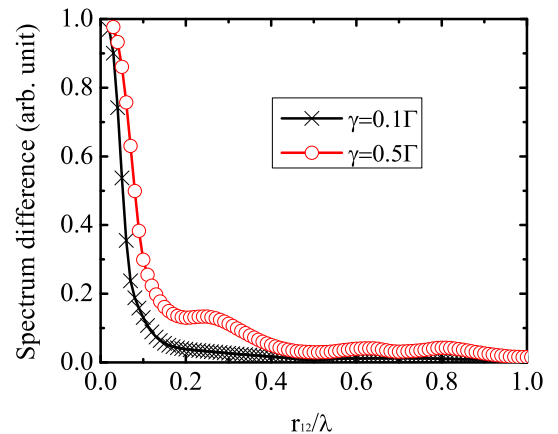}
\caption{(Color online) The normalized spectrum difference between the cases with and without $V_{dd}^{(nw)}$ for different emitter separations. The curve with cross symbol is the result when $\gamma=0.1\Gamma$, and the curve with circle symbol is the result when $\gamma=0.5\Gamma$. }
\end{figure}

In the last two subsections, we have seen that the dipole-dipole interaction induced by the non-guide vacuum modes can significantly affect the emitter dynamics and emission spectra when the emitter separation is much smaller than the resonance wavelength. To further quantify the effects of the dipole-dipole interaction induced by the non-guide vacuum modes, we define the normalized spectrum difference with and without $V_{dd}^{(nw)}$ as
\begin{equation}
\Delta_{SD} =\frac{1}{2}\Big\{\frac{\sum_{k}|I_{R}^{1}(k)-I_{R}^{2}(k)|}{\sum_{k}[I_{R}^{1}(k)+I_{R}^{2}(k)]}+\frac{\sum_{k}|I_{T}^{1}(k)-I_{T}^{2}(k)|}{\sum_{k}[2-I_{T}^{1}(k)-I_{T}^{2}(k)]}\Big\}
\end{equation}
where $I_{R}^{1}(k)$ ($I_{T}^{1}(k)$) is the reflection (transmission) spectrum with  $V_{dd}^{(nw)}$ and $I_{R}^{2}(k)$ ($I_{T}^{2}(k)$) is the reflection (transmission) spectrum without  $V_{dd}^{(nw)}$. Since the background transmission is 1, in the second term of Eq. (35) the quantity $1-I_{T}^{1,2}(k)$ is used instead of $I_{T}^{1,2}(k)$ to avoid divergence in the numerator. If the emission spectra with and without $V_{dd}^{(nw)}$ are completely identical, $\Delta_{SD}=0$. On the contrary, if the emission spectra with and without $V_{dd}^{(nw)}$ are completely different (i.e., have no any overlaps), $\Delta_{SD}=1$. Therefore, the quantity shown in Eq. (35) is a good measure of spectrum difference with and without $V_{dd}^{(nw)}$.  

In Fig. 4, we plot the spectrum difference for different emitter separations with two different $\gamma$ ($\gamma=0.1\Gamma$ and $\gamma=0.5\Gamma$). When the emitter separation $r_{12}$ approaches zero, the spectrum difference $\Delta_{SD}$ approaches 1 which means that the spectra with and without $V_{dd}^{(nw)}$ for small emitter separation are almost completely different. When the emitter separation $r_{12}$ is of the order of or larger than the resonant wavelength, $\Delta_{SD}$ is close to zero which indicates that the spectra with and without $V_{dd}^{(nw)}$ for large emitter separation are almost the same. These observations are consistent with the results shown in previous sections. 

For $\gamma=0.1\Gamma$, the spectrum difference is 0.5 when $r_{12}\simeq 0.05\lambda$. For $\gamma=0.5\Gamma$, the spectrum difference is 0.5 when $r_{12}\simeq 0.08\lambda$. For both cases, the spectrum difference is 0.5 when $|V_{dd}^{(nw)}|\approx 2\Gamma$. When $|V_{dd}^{(nw)}|\approx 0.2\Gamma$, i.e., $r_{12}\simeq 0.1\lambda$ for $\gamma=0.1\Gamma$ and $r_{12}\simeq 0.3\lambda$ for $\gamma=0.5\Gamma$, the spectrum difference is about $10\%$. Therefore, when $|V_{dd}^{(nw)}|< 0.2\Gamma$, we can safely neglect the effect of $V_{dd}^{(nw)}$. Otherwise, the effect of $V_{dd}^{(nw)}$ should be taken into account.

\begin{figure*}
\includegraphics[width=0.65\columnwidth, bb=0 100 575 575]{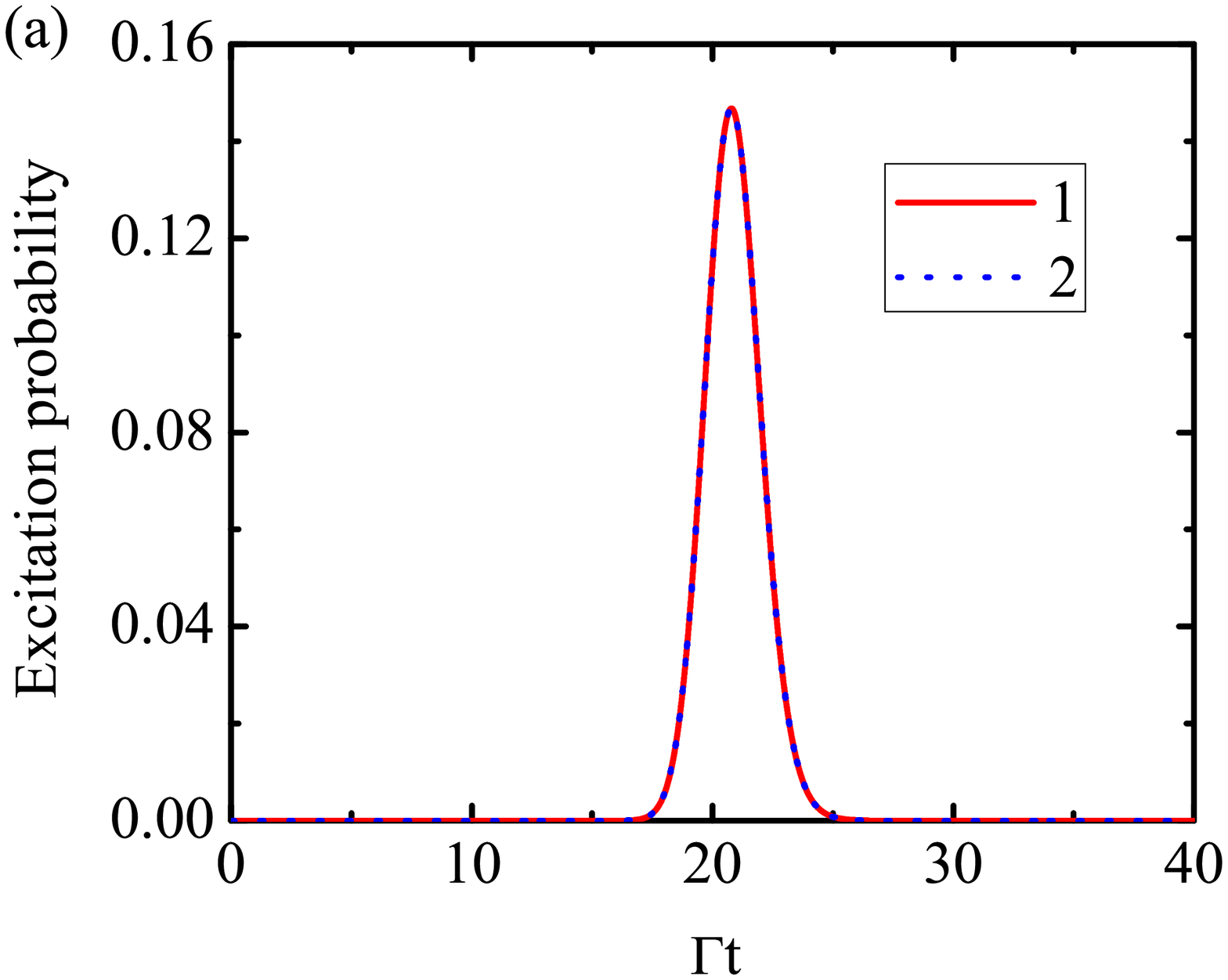}
\includegraphics[width=0.65\columnwidth, bb=0 100 575 575]{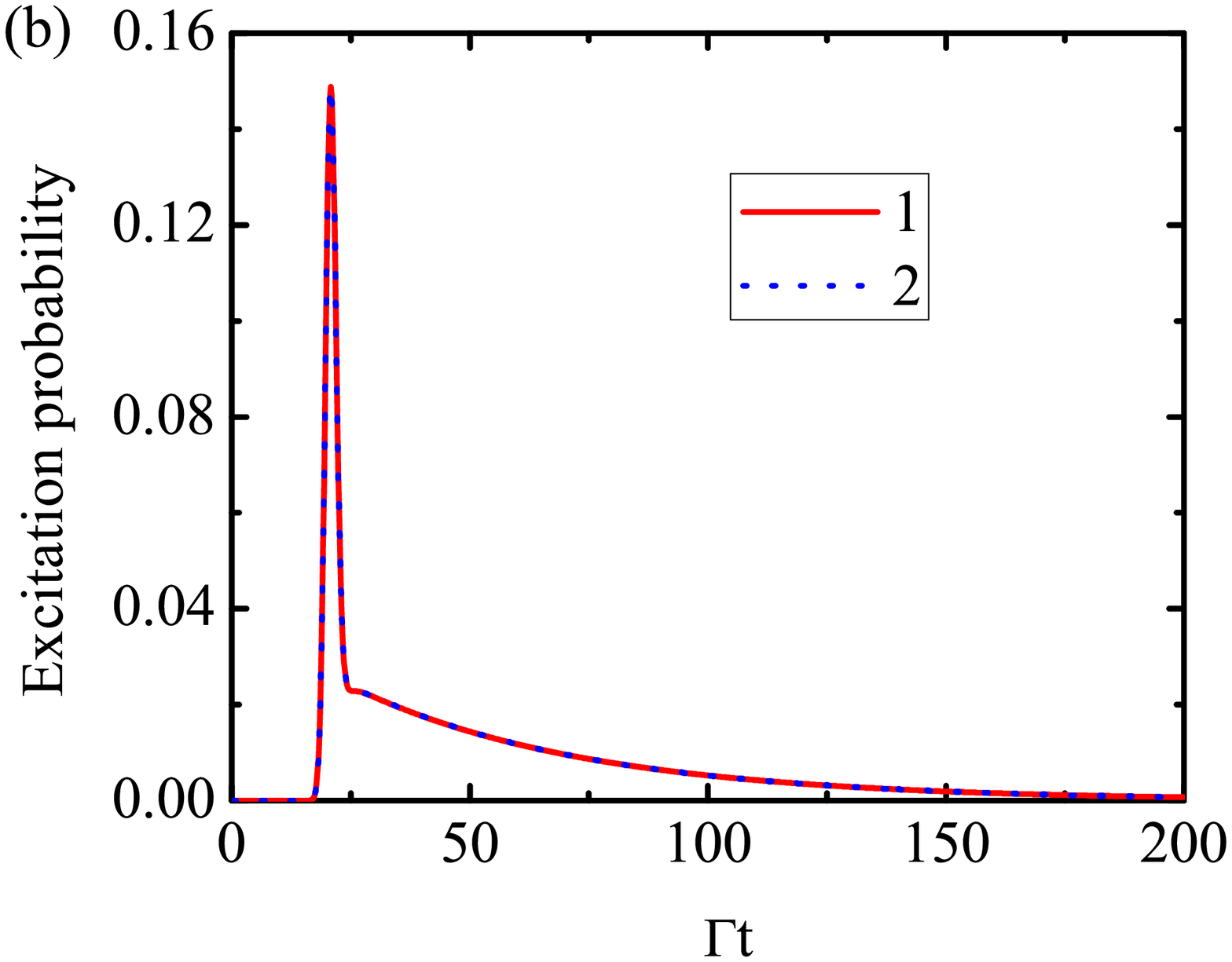}
\includegraphics[width=0.65\columnwidth, bb=0 100 575 575]{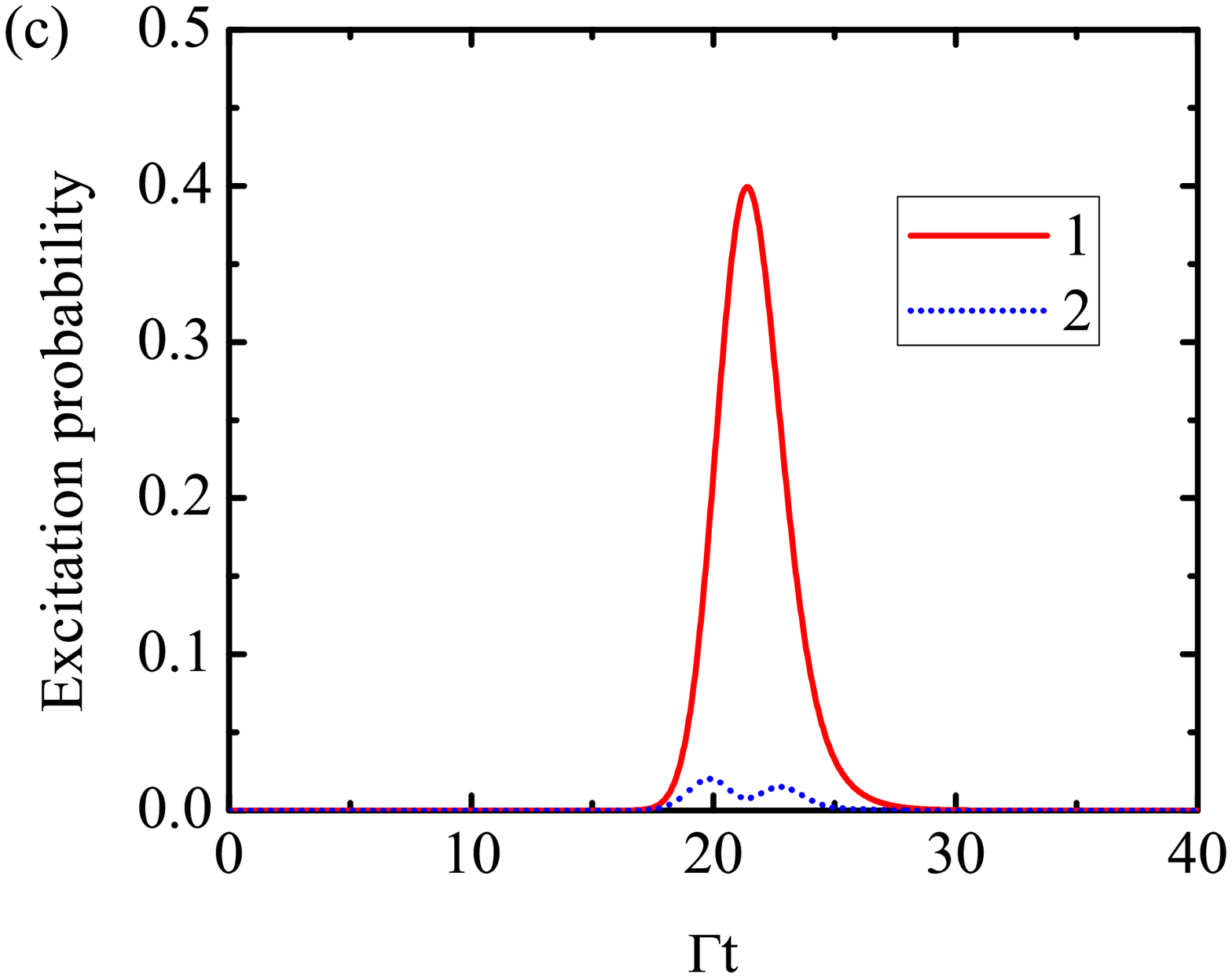}
\includegraphics[width=0.65\columnwidth, bb=0 100 575 575]{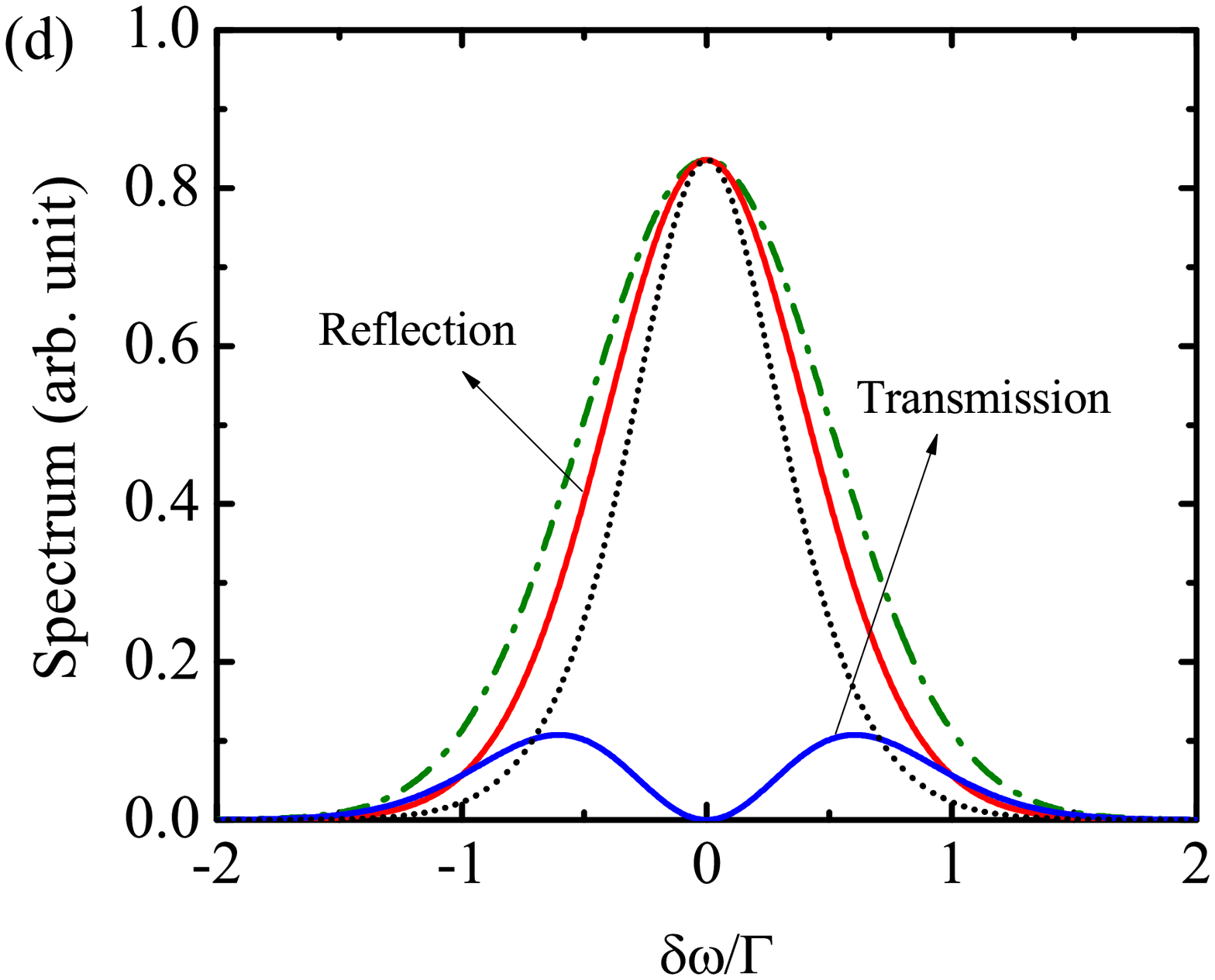}
\includegraphics[width=0.65\columnwidth, bb=0 100 575 575]{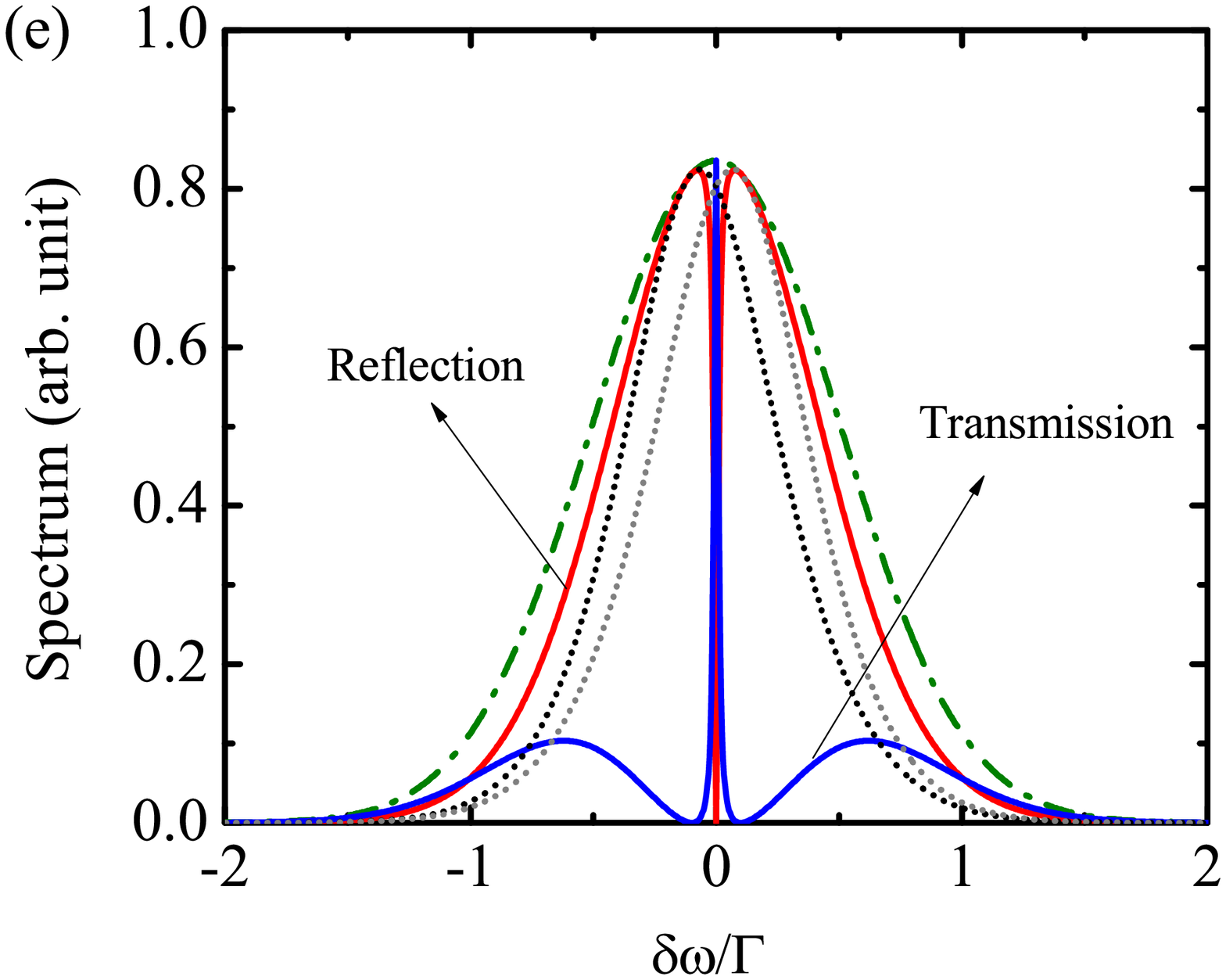}
\includegraphics[width=0.65\columnwidth, bb=0 100 575 575]{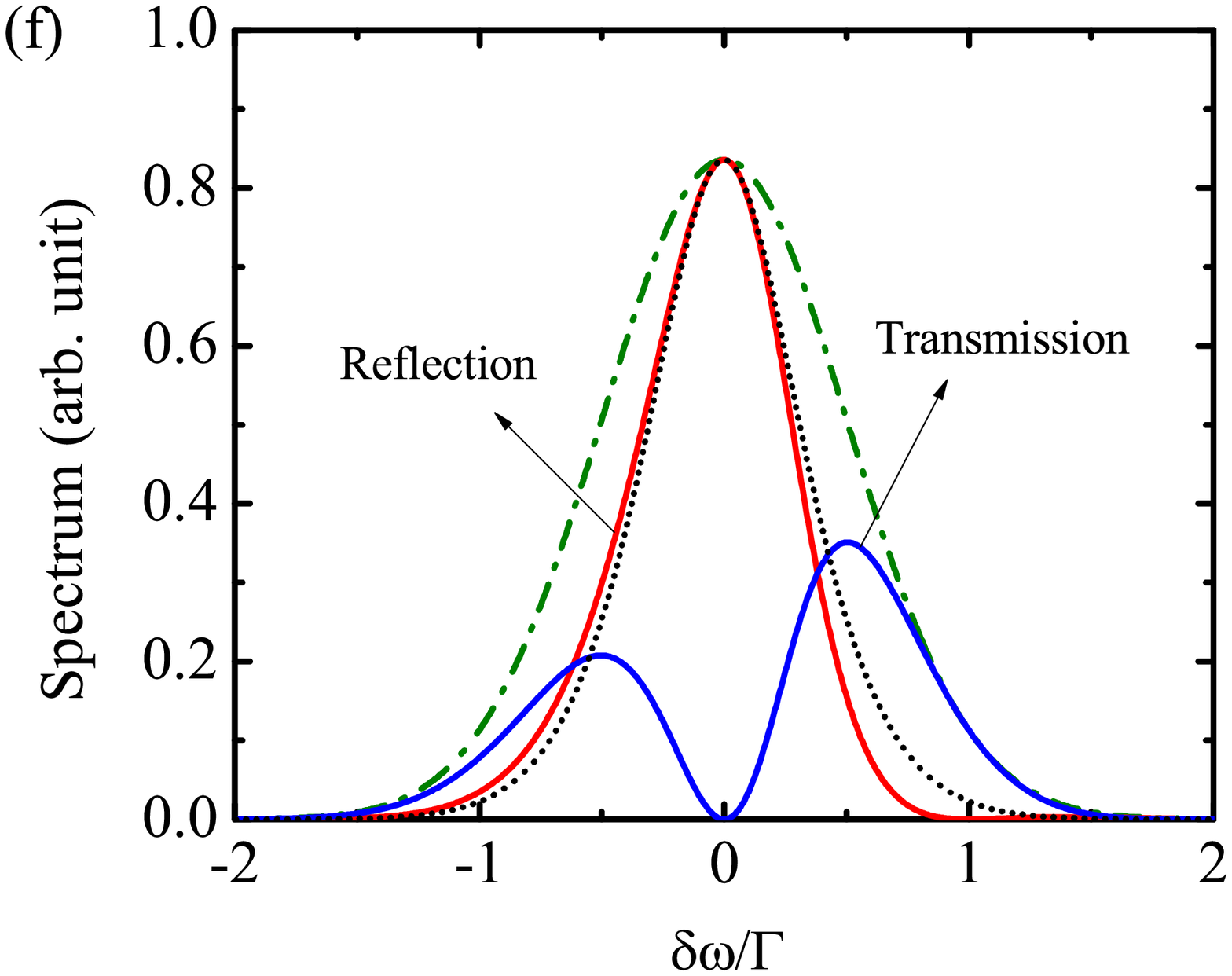}
\caption{(Color online) (a-c) Emitter excitation probabilities as a function of time for different emitter energy gaps with single Gaussian photon pulse input. The red solid line is the excitation for emitter 1, and the blue dotted line is the excitation for emitter 2. (d-f) Emission spectra for different emitter energy gaps. The green dahsed-dotted line is the input photon spectrum, the red and blue solid lines are the reflection and transmission photon spectra of the two-emitter system, the black and gray dotted lines are the reflection spectrum of independent emitter case.   Parameters: $\Delta_{0}v_{g}=\Gamma, \gamma=0$. $z_{1}=20/\Delta_{0}, z_{12}=0.5\lambda$. (a, d) $\Delta \omega_{12}=0$. (b, e) $\Delta \omega_{12}=0.2\Gamma$. (c, f) $\Delta \omega_{12}=2\Gamma$. }
\end{figure*}

\section{non-identical emitters}

In this section, we consider two-emitter case to illustrate the effects of non-identical emitters. For this purpose, we let their transition frequencies be not the same, i.e.,  $\omega_{a}^{j}\neq \omega_{a}^{l}$ or $k_{j}\neq k_{l}$ if $j\neq l$. 

For a two-emitter system, the excitation dynamics of the emitters are given by 
\begin{align}
\dot{\alpha}_{1}(t)&=b_{1}(t)-V_{11}\alpha_{1}(t)-V_{12}e^{ik_{a}r_{12}}e^{i\Delta\omega_{12}t}\alpha_{2}(t-\frac{z_{12}}{v_{g}}), \\
\dot{\alpha}_{2}(t)&=b_{2}(t)-V_{22}\alpha_{2}(t)-V_{21}e^{ik_{a}r_{12}}e^{i\Delta\omega_{21}t}\alpha_{2}(t-\frac{z_{12}}{v_{g}}), 
\end{align}
where $\Delta\omega_{ij}=\omega_{i}-\omega_{j}$ with $i,j=1,2$. The coupling matrix between these two single-emitter excited states reads
\begin{equation}
V(t)=\begin{bmatrix} V_{11} & V_{12}e^{ik_a z_{12}}e^{i\Delta\omega_{12}t} \\ V_{12}e^{ik_a z_{12}}e^{-i\Delta\omega_{12}t} & V_{22} \end{bmatrix},
\end{equation}
which is time-dependent. It is seen that the coupling between the two emitters is modulated by the energy difference between these two emitters.  When $\Delta\omega_{12}=0$, it reduces to the case of identical emitters. When $\Delta\omega_{12}=\infty$, the rapid oscillations can erase the off-diagonal terms in Eq. (39) and thus eliminate the coupling between the two emitters.  The instantaneous single-emitter excited eigenstates are $|\psi_{\pm}(t)\rangle=|eg\rangle \pm e^{-i\Delta\omega_{12} t}|ge\rangle$ and their corresponding eigenvalues are $E_{\pm}=V_{11}\pm V_{12}e^{ik_a r_{12}}$. Although the eigenvalues are the same as those with identical emitters, the eigenstates here are time-dependent which are quite different from those with identical emitters. Due to the time modulation factor, the two states $|\psi_{+}\rangle$ state and $|\psi_{-}\rangle$ can interchange to each other as time evolves.

For the two-emitter system, $M(\delta k)$ in Eqs. (23) and (24) is given by
\begin{equation}
M(\delta k)=\begin{bmatrix} V_{11}-i(\delta k-\Delta k_{1})v_{g} & V_{12}e^{ik z_{12}} \\ V_{21}e^{ik z_{12}} & V_{22}-i(\delta k-\Delta k_{2})v_{g} \end{bmatrix}
\end{equation}
For a single incident photon pulse, we obtain the reflection and transmission photon spectra given by
\begin{widetext}
\begin{align}
\beta_{R}(\delta k)&=\frac{\Gamma}{2}e^{2ikr_{1}}\beta_{0}(\delta k)\frac{2M_{12}(\delta k)e^{ikz_{12}}-M_{11}(\delta k)e^{2ikz_{12}}-M_{22}(\delta k)}{M_{11}(\delta k)M_{22}(\delta k)-M_{12}^{2}(\delta k)}, \\
\beta_{T}(\delta k)&=\beta_{0}(\delta k)\Big\{1-\frac{\Gamma}{2}\frac{M_{11}(\delta k)+M_{22}(\delta k)-2M_{12}(\delta k)\text{cos}(kz_{12})}{M_{11}(\delta k)M_{22}(\delta k)-M_{12}^{2}(\delta k)} \Big\},
\end{align}
\end{widetext}
It is seen that $\beta_{R}(\delta k)$ and  $\beta_{T}(\delta k)$ depends on $\beta_{0}(\delta k)$ but not other frequency components. Therefore, no frequency conversion can occur here.

\subsection{Without non-waveguide modes} 

In this subsection, we first consider the case without non-waveguide modes, i.e., $\gamma=0$. In this case, $V_{11}=V_{22}=V_{12}=\Gamma/2$.
Here, we compare the excitation dynamics (Fig. 5(a-c)) and emission photon spectra (Fig. 5(d-f)) when $r_{12}=0.5\lambda$ for three different emitter energy differences, i.e.,  $\Delta \omega_{12}=0, 0.2\Gamma$, and $2\Gamma$. In these numerical examples, we assume that the input single photon pulse has a Gaussian shape as shown in Eq. (33) with $\Delta_0=\Gamma$. For $r_{12}=0.5\lambda$, $V_{12}^{(w)}e^{ik_{a}r_{12}}=-0.5\Gamma$. The energy shift by the dipole-dipole interaction induced by the waveguide photon modes is zero, and the decay rates for the two single-emitter excited eigenstates ($|+\rangle$ and $|-\rangle$) are given by $\Gamma$ and $0$ with one being superradiant and the other being subradiant. 

When the two emitters are identical ($\Delta \omega_{12}=0$), both emitters are excited and then deexcited together as the incident pulse propagates through (Fig. 5(a)). From Eqs. (40) and (41), it is readily seen that $\beta_{R}(0)=-e^{2ikr_{1}}\beta_{0}(\delta k)$ and $\beta_{T}(0)=0$, i.e., the resonant frequency is completely reflected. Although independent-emitter model can also explain the total reflection of the resonance frequency (black dotted line), it cannot explain the broader reflection linewidth for the two-emitter system (red solid line). The broader linewidth is the signature of the superradiant state induced by the collective interaction between the two emitters (Fig. 5(d)).  

The results when the two emitters have close but non-zero energy difference (e.g., $\Delta \omega_{12}=0.2\Gamma$ with $\Delta\omega_{a}^{(1,2)}=\pm 0.1\Gamma$) are shown in Figs. 5(b) and 5(e). From Fig. 5(b), we see that the two emitters are also excited and then deexcited together. However, different from the case of identical emitters (Fig. 5(a)), the emitter excitations when $\Delta \omega_{12}=0.2\Gamma$ can last much longer (Fig. 5(b)). This indicates that the subradiant state can be populated when there is a small energy difference between the two emitters. This can be explained by the fact that the superradiant and subradiant states can interchange to each other when there is a time modulation factor in the coupling matrix as shown in Eq. (38). In contrast, for two identical emitters, the subradiant state will be never populated when $r_{12}=0.5\lambda$. The emission photon spectra also become very distinctive (Fig. 5(e)). Instead of being completely reflected at the resonant frequency as in the identical emitter case, a very narrow transmission window appears around the resonance frequency when the two emitters has close but non-zero energy difference. This transparency can be seen from Eqs. (40) and (41). For $\delta k r_{12}\ll 1$, we can see from Eq. (40) that when $\delta k v_{g}=(\Delta \omega_{1}+\Delta \omega_{2})/2$, we have $\beta_{R}(0)=0$ which means the resonance frequency can be completely transparent. However, if we neglect the dipole-dipole coupling between the two emitters ($V_{12}=0$), we have $|\beta_{R}(0)/\beta_{0}(0)|=1/[1+(2\Delta\omega_{12}/\Gamma)^{2}]$ which is close to 1 when $\Delta\omega_{12}\ll \Gamma$. Therefore, the dipole-dipole interaction here is critical for the transmission of the resonance frequency and the phenomena here can be also called as ``didople-dipole induced eletromagnetic transparency (DIET)". Actually, the transparency is the result of destructive interference between two emission channels. The DIET has been studied in an atomic ensemble where semiclassical and mean-field theory are applied \cite{Joseph163603}. Here we provide an ab initio calculation for this phenomenon and the system here can be easier to realize in experiment. The DIET here may be used as single photon switch by tuning the emitter energy.  

\begin{figure*}
\includegraphics[width=0.65\columnwidth, bb=0 100 575 575]{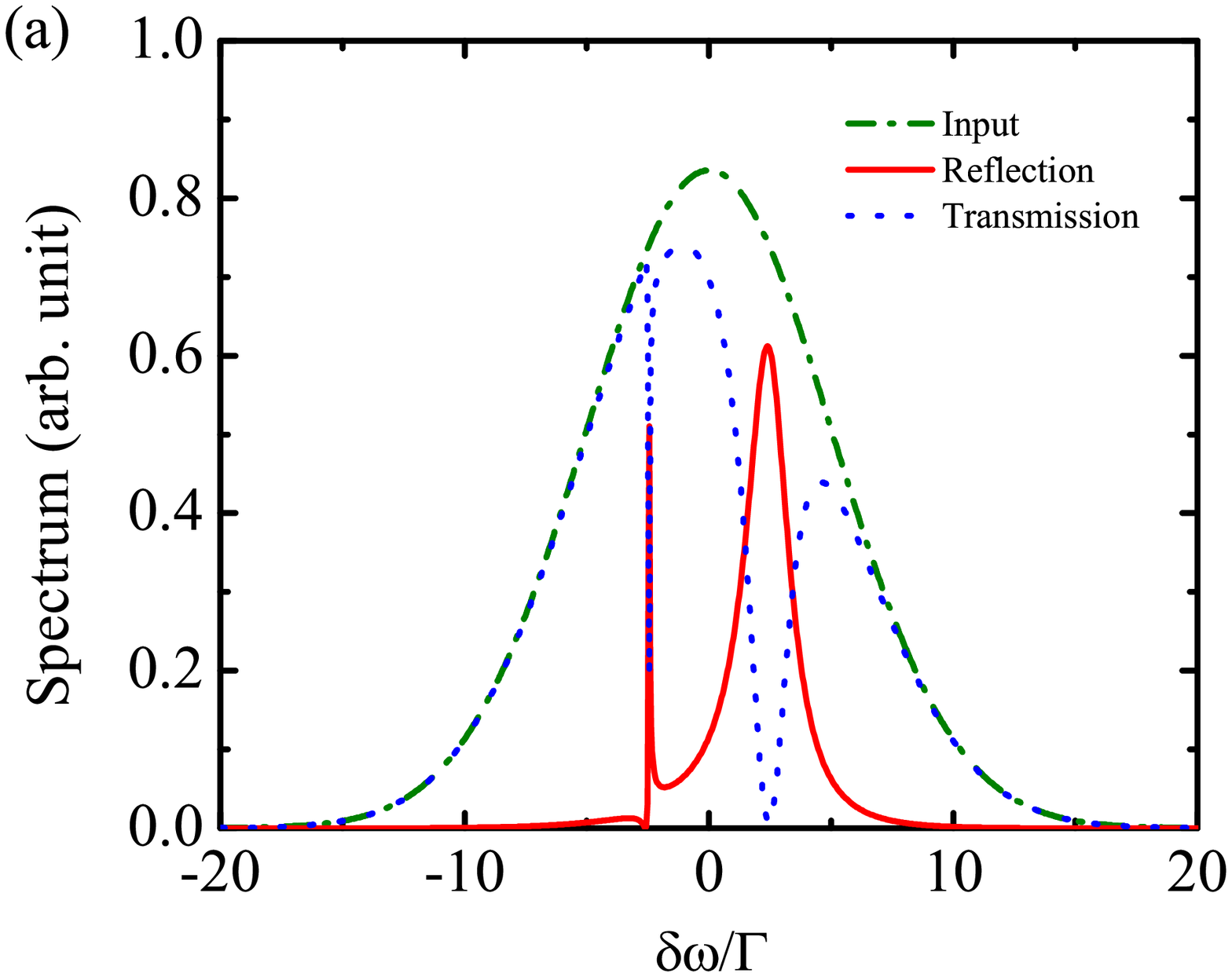}
\includegraphics[width=0.65\columnwidth, bb=0 100 575 575]{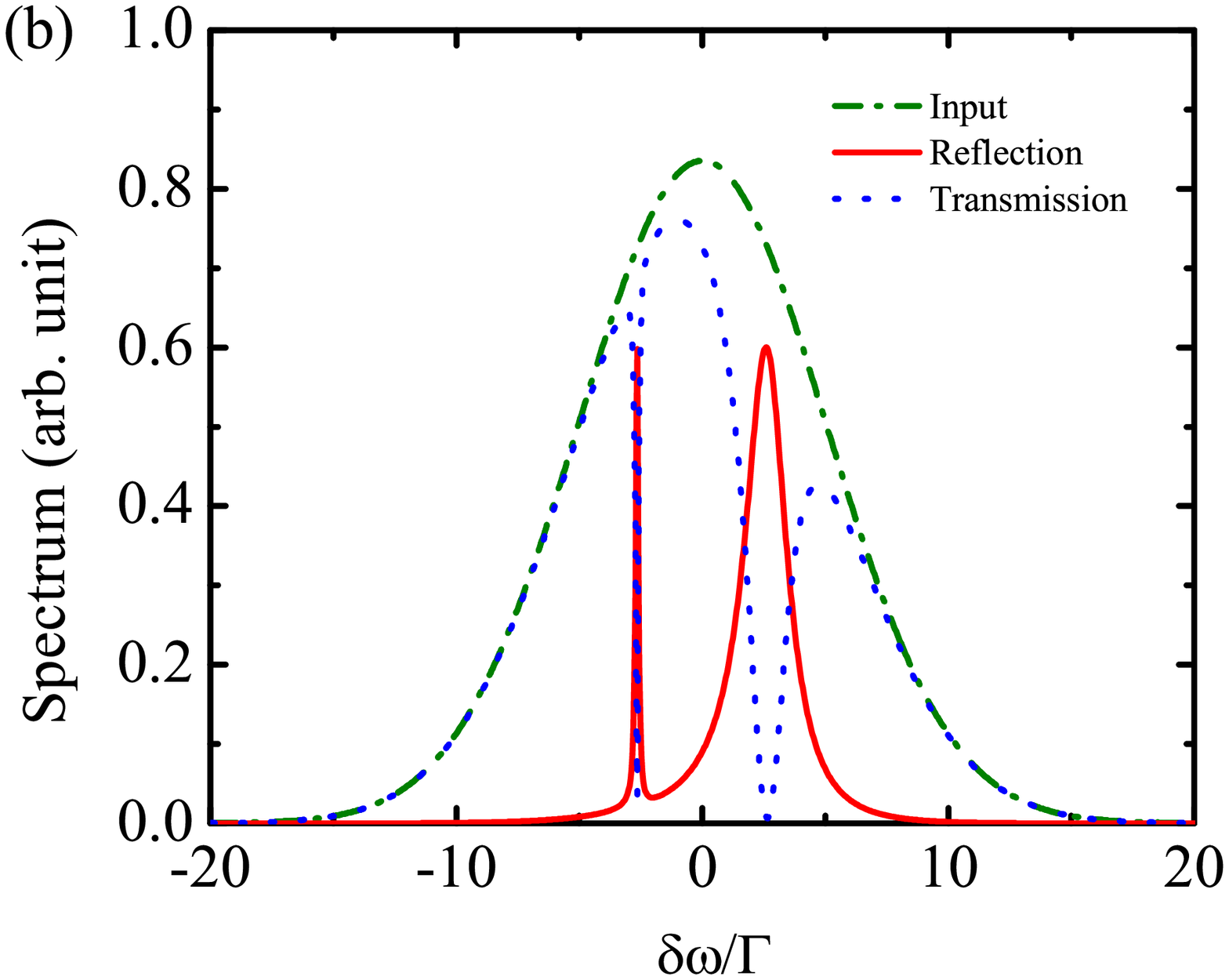}
\includegraphics[width=0.65\columnwidth, bb=0 100 575 575]{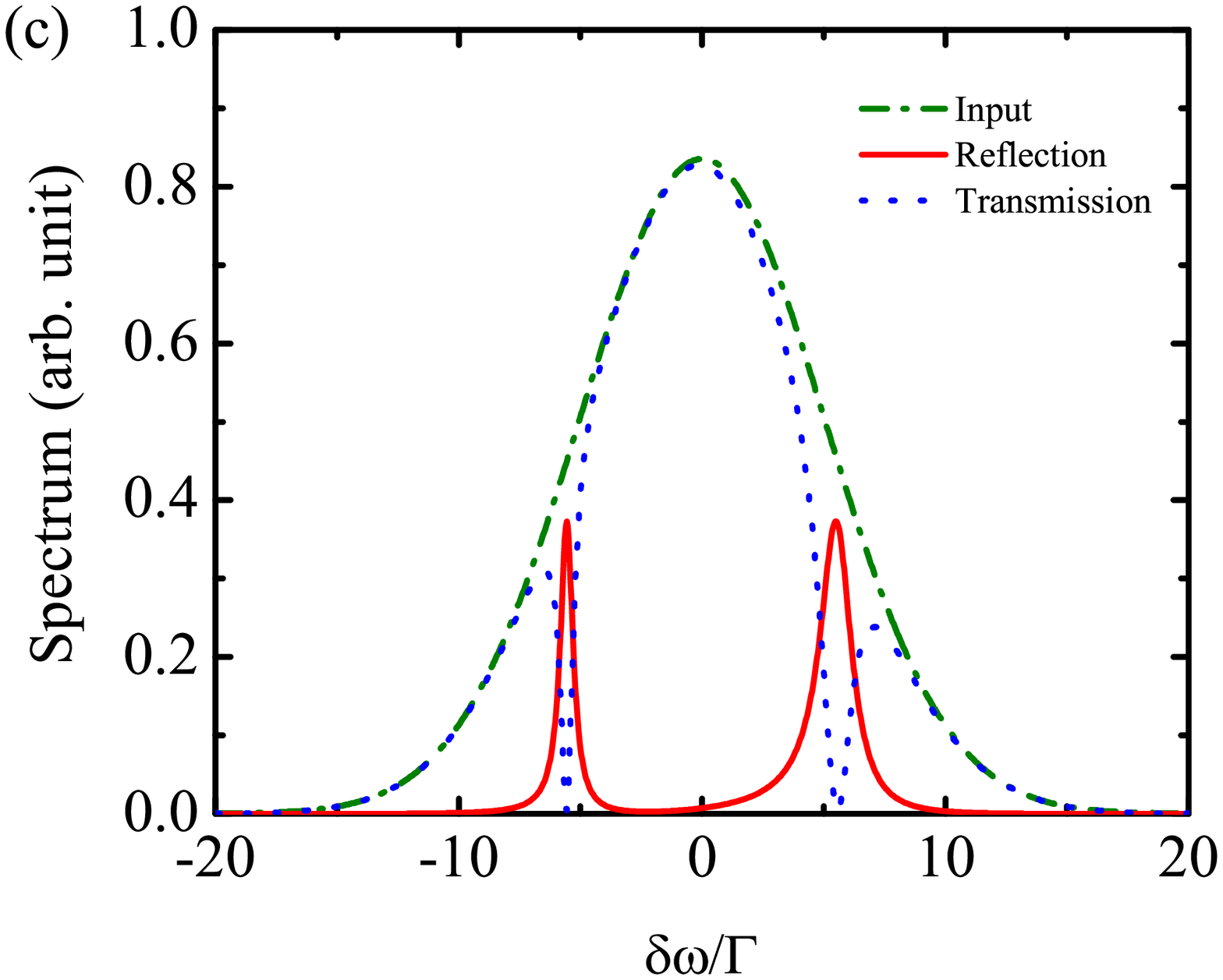}
\caption{(Color online) Emission photon spectrum for different emitter energy gaps including the effects of non-waveguide modes. The green dahsed-dotted line is the input photon spectrum, the red (blue) solid lines are the reflection (transmission) photon spectra.   Parameters: $\Delta_{0}v_{g}=10\Gamma, \gamma=0.1\Gamma$. $z_{1}=20/\Delta_{0}, z_{12}=0.05\lambda$. (a) $\Delta \omega_{12}=0$. (b) $\Delta \omega_{12}=2\Gamma$. (c) $\Delta \omega_{12}=10\Gamma$. }
\end{figure*}

When the energy difference of the two-emitters is large, e.g. $2\Gamma$ and one emitter has transition frequency resonant with the center frequency of the incident photon, we can see that one emitter is excited as a single emitter case, but the other one is rarely excited by the input photon pulse due to large detuning (Fig. 5(c)). The emission spectra are also similar to those of the independent emitter case. Therefore, when the two emitters has a large energy difference (i.e., much greater than their dipole-dipole interaction energy), they behave as independent emitters.

\begin{figure*}
\includegraphics[width=0.9\columnwidth, bb=0 100 575 575]{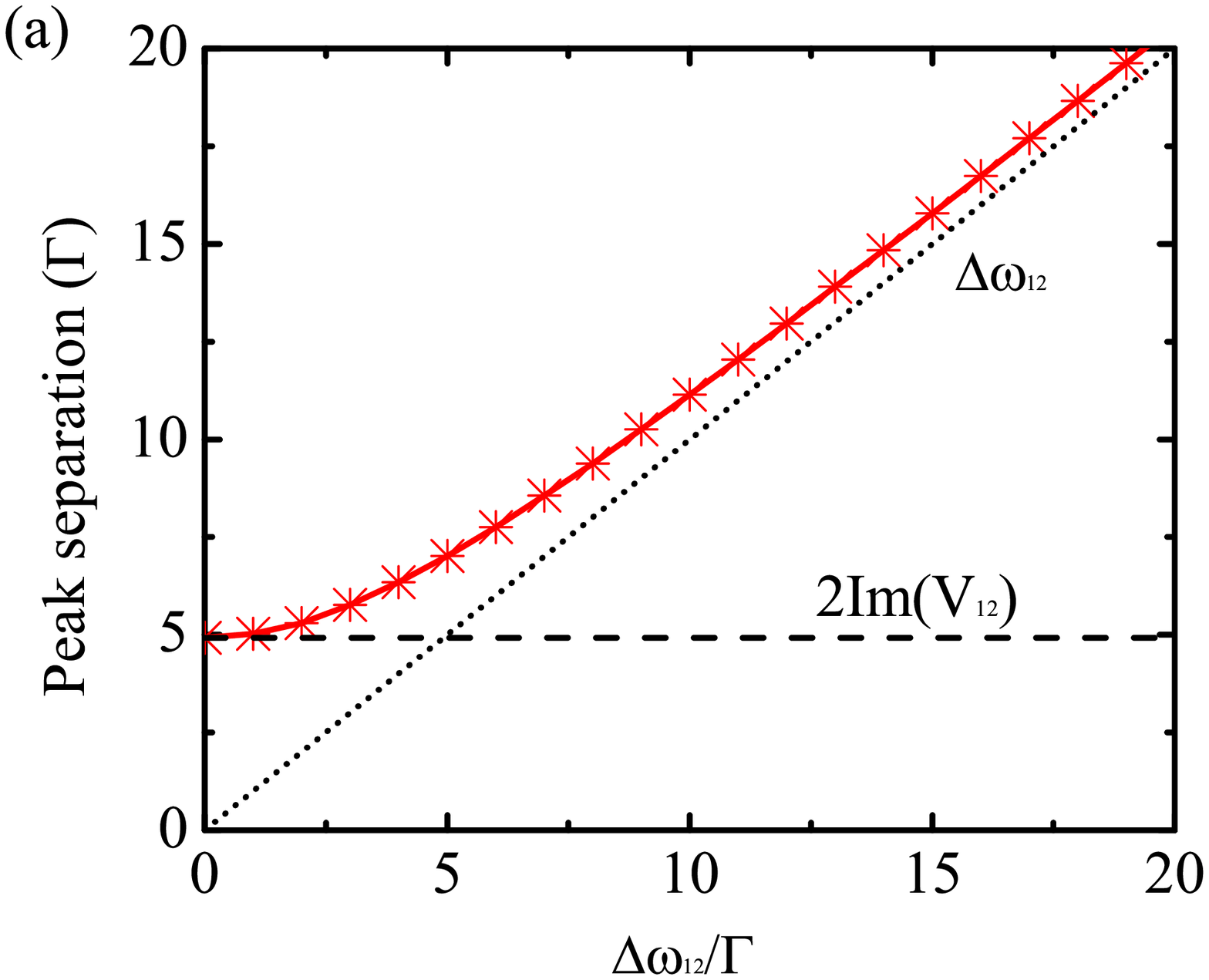}
\includegraphics[width=0.9\columnwidth, bb=0 100 575 575]{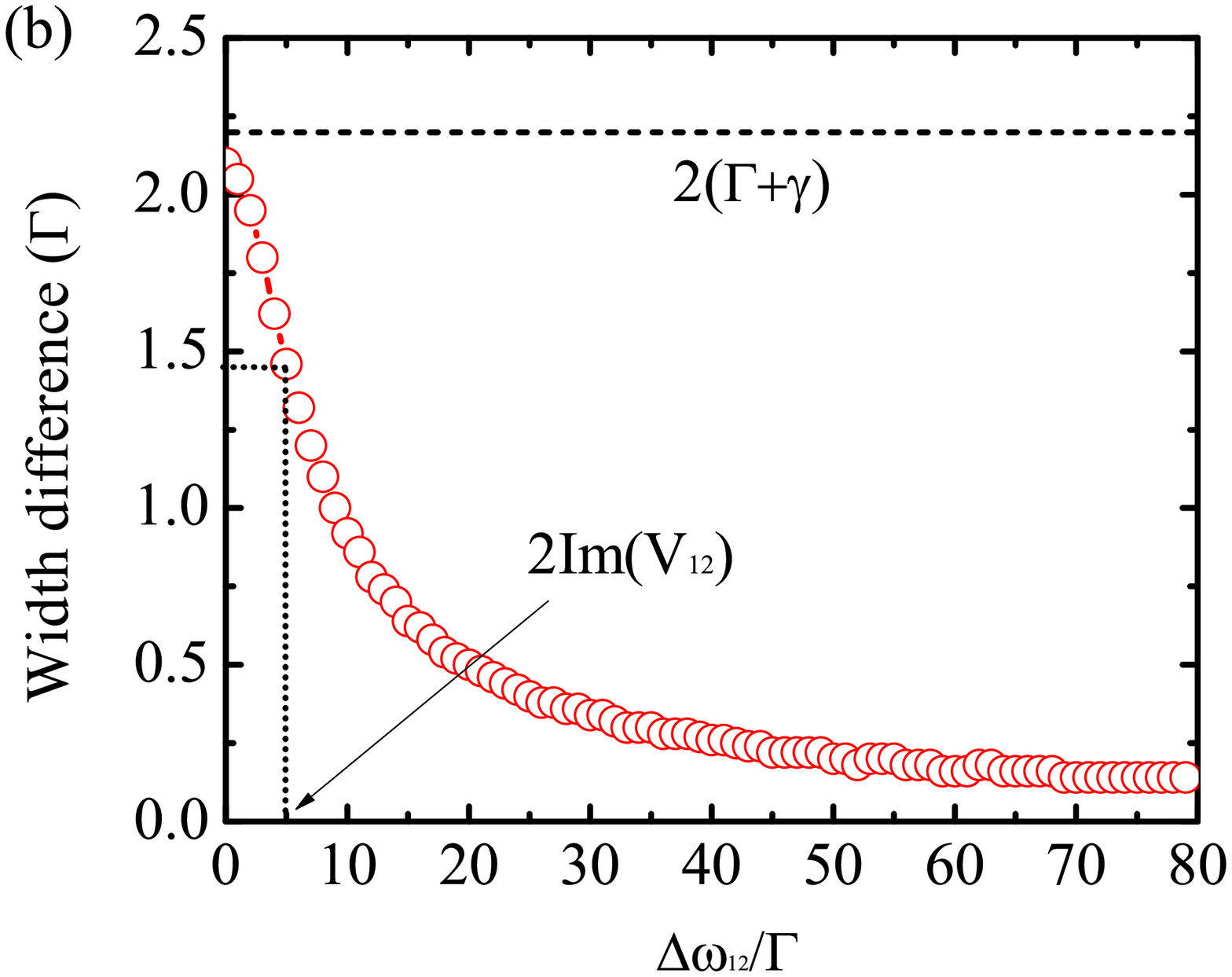}
\caption{(Color online) Transition from coupled emitters to independent emitters. (a) The reflection peak separation as a function of emitter energy difference. (b) The linewidth difference between the two reflection peaks as a function of emitter energy difference. Parameters: $\gamma=0.1\Gamma$, and $z_{12}=0.05\lambda$.  }
\end{figure*}

\subsection{With non-waveguide modes}

In this subsection, we study how the emission spectra change when the emitter energy difference increases in the present of non-waveguide modes, i.e. $\gamma\neq 0$. 
The numerical results when $a=0.05\lambda$ and $\gamma=0.1\Gamma$ are shown in Fig. 6. In this case, the dipole-dipole interaction is $V_{12}e^{ik_{a}r_{12}}=0.52\Gamma+i2.46\Gamma$.
The energy shifts due to the dipole-dipole interaction are $\pm 2.46\Gamma$, and the decay rates of the two eigenstates are $\Gamma_{+}=1.07\Gamma$ and $\Gamma_{-}=0.03\Gamma$, respectively.
In the numerical results, we assume that the incident photon pulse has a Gaussian shape with $\Delta_{0}=10\Gamma$.

The emission spectra when the two emitters are identical are shown in Fig. 6(a). There are two reflection peaks at around $\pm 2.46\Gamma$ with one being very broad and the other being very sharp. The peak positions are the same as the energy shifts due to the dipole-dipole interaction. The broad peak has a width of about $2.13\Gamma$ which is $2\Gamma_{+}$ due to the reflection from the superradiant state. The sharp peak has a width of about $0.06\Gamma$ which is $2\Gamma_{-}$  due to the reflection from the subradiant eigenstate.  

When the two emitters have different transition frequencies, for example $\Delta \omega_{12}=2\Gamma$, there are also two reflection peaks with one being the superradiant peak and the other one being the subradiant peak (Fig. 6(b)). The positions of the peaks are about $\pm 2.61\Gamma$  which are slightly larger than the energy shifts due to the dipole-dipole interaction. The superradiant peak has a width of about $2.04\Gamma$ which is slightly narrower than that of the identical emitters, and the subradiant peak has a width of about $0.12\Gamma$ which is slightly broader than that of the identical emitters. Although the difference between these two peaks decreases, the dipole-dipole interaction still plays an important role when the energy difference is of the order of the dipole-dipole induced energy shift. 

If we continue to increase the energy difference such that the energy difference between the two emitters is much larger than the dipole-dipole induced energy shift, for example $\Delta \omega_{12}=10\Gamma$, the emission spectra are quite different from those in Fig. 6(a) and 6(b). The two reflection peaks become more similar to each other with one peak having a width of about $1.56\Gamma$ and the other one having a width of about $0.66\Gamma$. The positions of the two peaks are about $\pm 5.5\Gamma$ which is quite different from the energy shifts due to the dipole-dipole interaction. The separation between the two peaks is about $11\Gamma$ which is close to the energy difference of the two emitters which indicates that they behave more like independent emitters.

\subsection{Transition from coupled emitters to independent emitters}

In the previous subsection, we have shown that the effective coupling between the emitters depends on the emitter energy difference. In this subsection, we quantify this dependence by calculating the peak separation and the linewidth difference of the two reflection peaks as a function of emitter energy difference. The peak separation as a function of emitter energy difference when $\gamma=0.1\Gamma$ and $r_{12}=0.05\lambda$ are shown in Fig. 7(a). The peak separation increases monotonically as the energy difference increases. When the two emitters are identical, i.e., $\Delta \omega_{12}=0$, the peak separation is equal to $2\text{Im}(V_{12})$ which means that the two emitters are strongly coupled to each other via the dipole-dipole interaction.  However, when the two emitters have a large energy difference, e.g. $\Delta \omega_{12}=20\Gamma$, the peak separation are close to $\Delta \omega_{12}$ which indicates that the two emitters behave mostly as independent emitters. Thus, the emitters can transit from coupled emitters to independent emitters by increasing the emitter energy difference. When $\Delta \omega_{12}<2\text{Im}(V_{12})$ or $\Delta \omega_{12}\sim 2\text{Im}(V_{12})$ the emitters can strongly couple to each other, but when$\Delta \omega_{12}\gg 2\text{Im}(V_{12})$ the emitters can be treated as independent emitters. 

In addition to the peak separation, we also study the linewidth difference between the two reflection peaks as a function of emitter energy difference which is shown in Fig. 7(b). When $\Delta \omega_{12}=0$, one reflection peak is a superradiant peak while the other one is a subradiant peak and their linewidth difference is about $2.1\Gamma$ which is close to the maximum value $2(\Gamma+\gamma)$. This means that when $\Delta \omega_{12}=0$, the collective effect plays an important role. However, when $\Delta \omega_{12}$ is large, the linewidth difference between the two reflection peaks approach zero which means that they behave like  independent emitters.  When $\Delta \omega_{12}=2\text{Im}(V_{12})$, the linewidth difference is about $66\%$ of the maximum linewidth difference.

\section{Beyond two emitters}

\begin{figure*}
\includegraphics[width=0.9\columnwidth, bb=0 100 575 575]{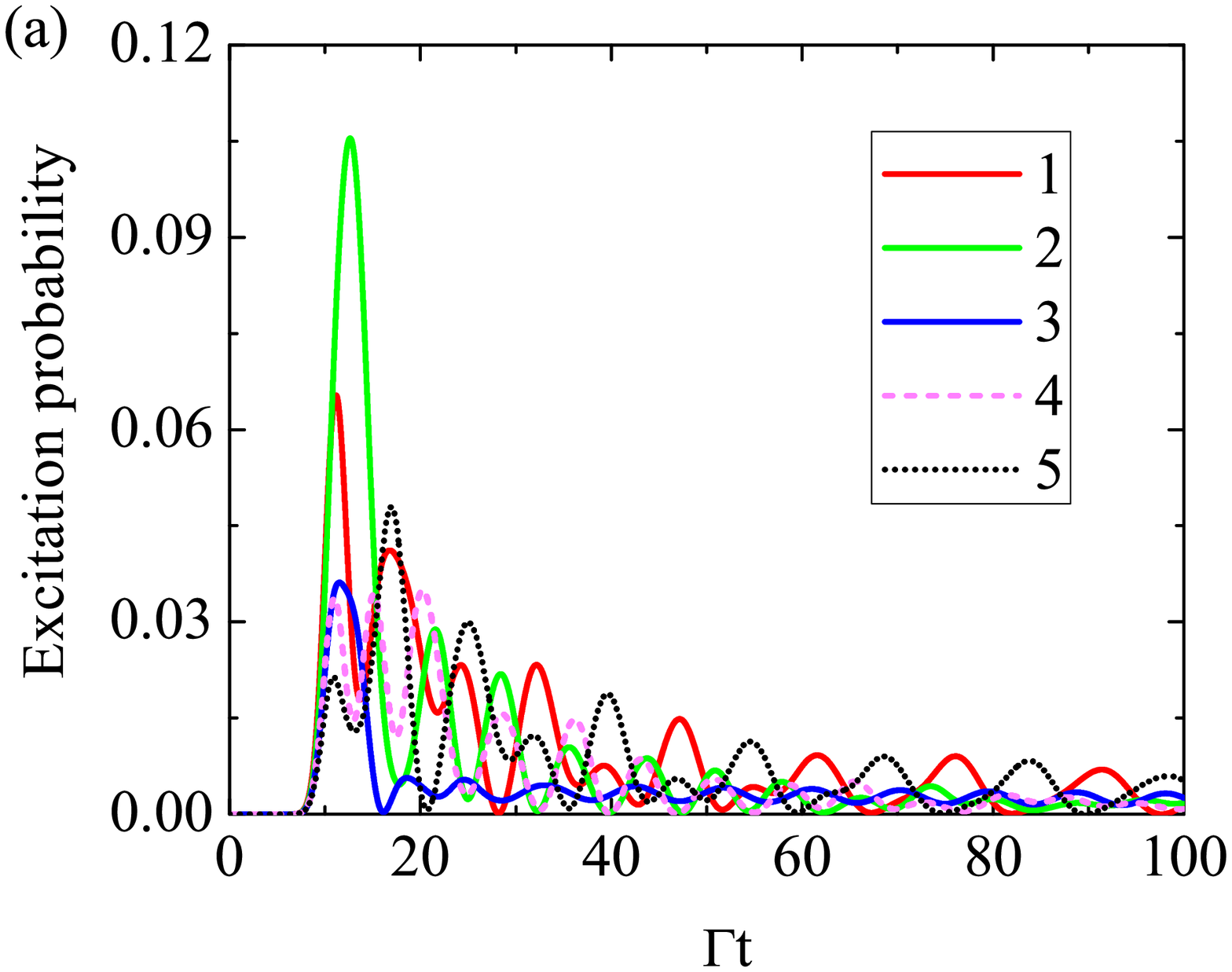}
\includegraphics[width=0.9\columnwidth, bb=0 100 575 575]{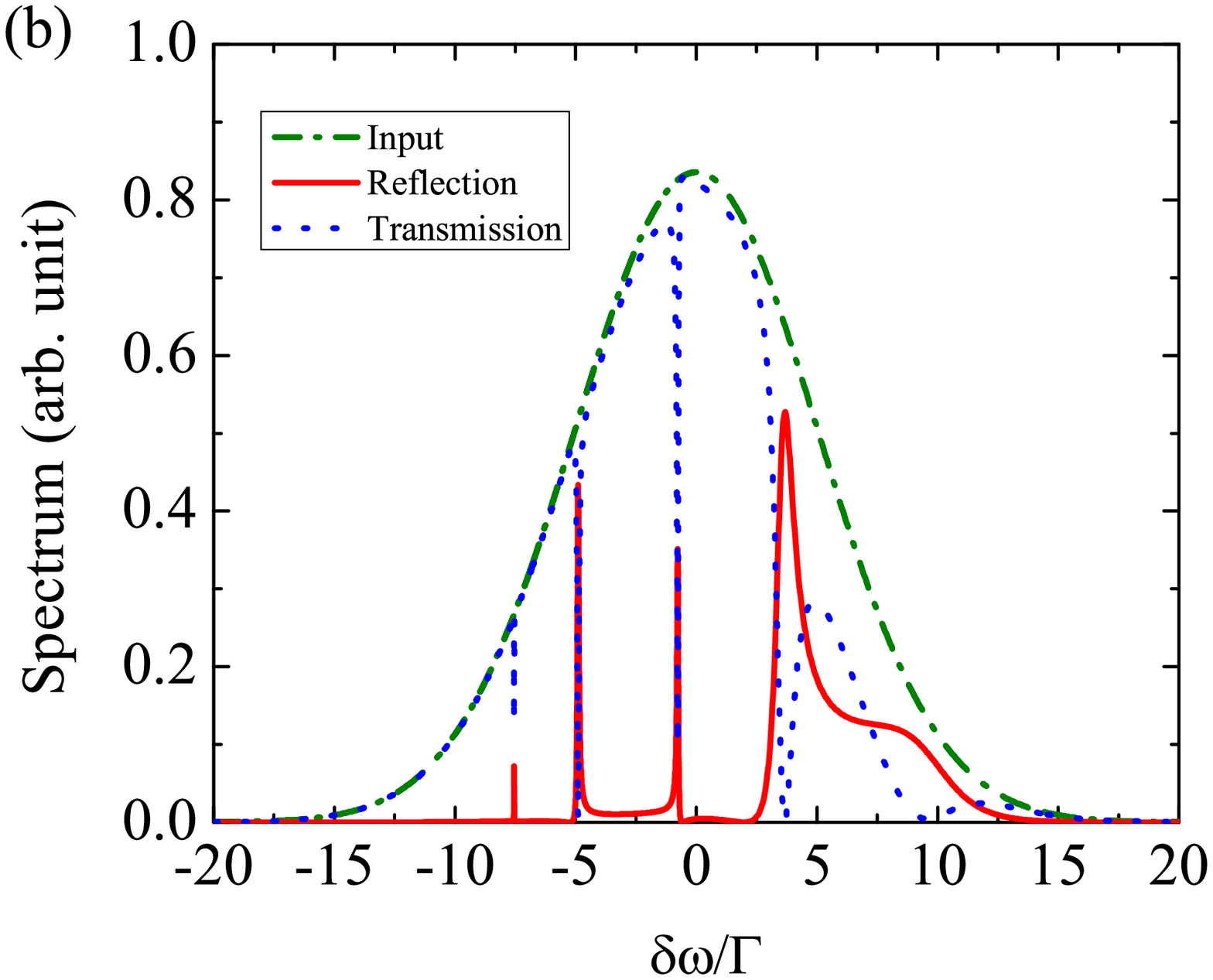}
\includegraphics[width=0.9\columnwidth, bb=0 100 575 575]{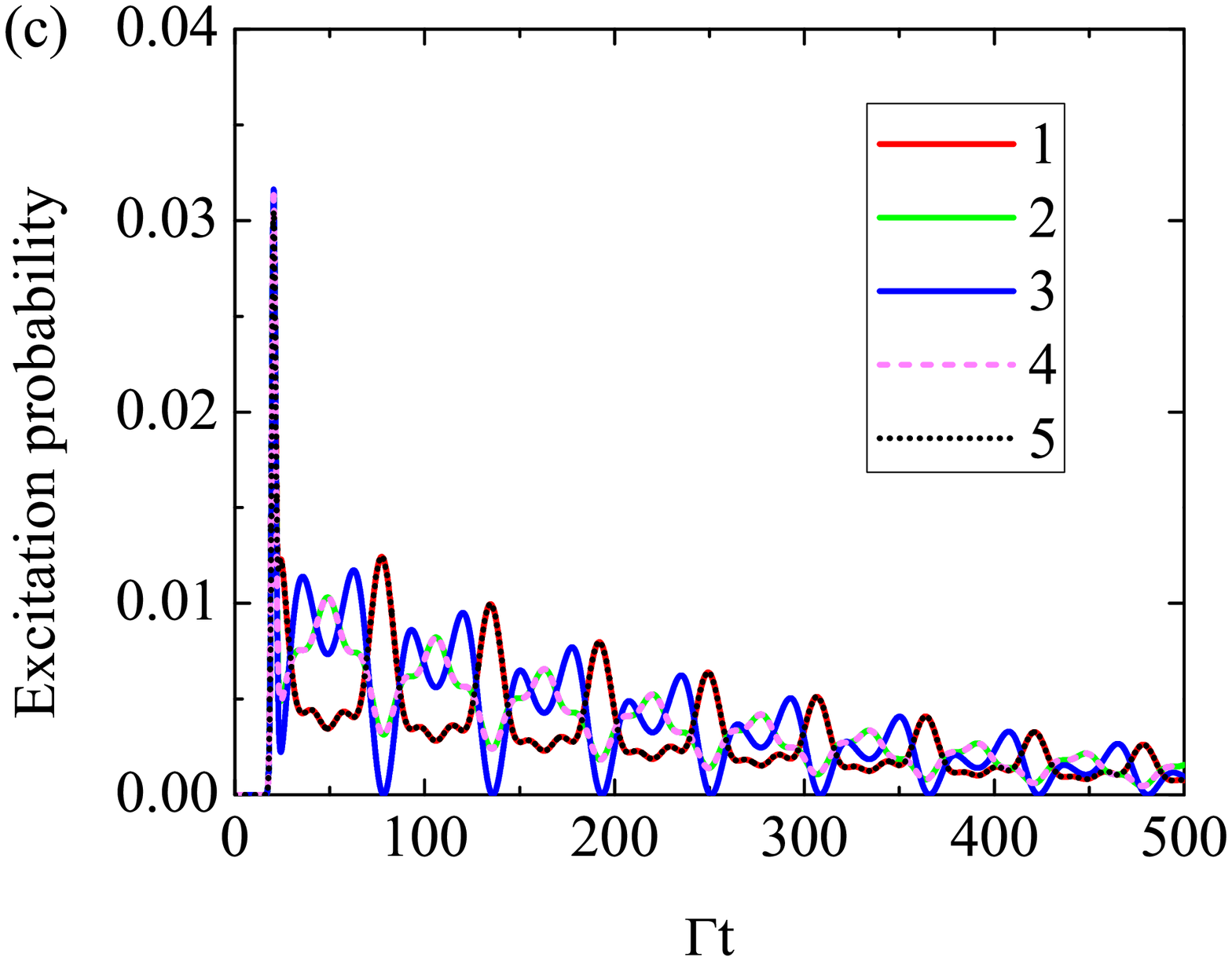}
\includegraphics[width=0.9\columnwidth, bb=0 100 575 575]{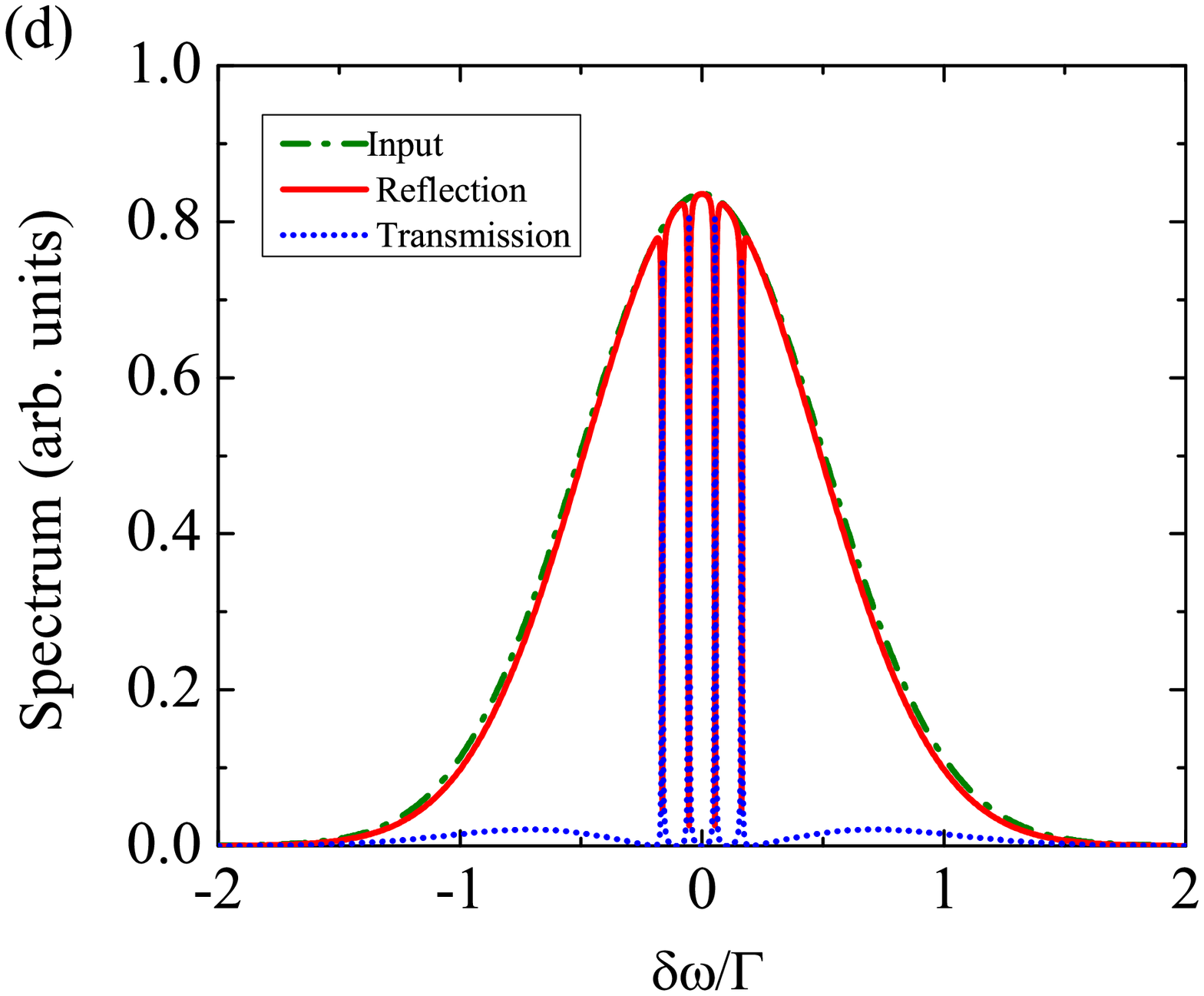}
\caption{(Color online) (a, c) Emitter excitation probabilities as a function of time. (b, d) Emission photon spectrum. (a, b) The emitters are identical with $z_{j,j+1}=0.05\lambda$, $\gamma=0.2\Gamma$, and $\Gamma_{0}=10\Gamma$. (c, d) The emitters are non-identical with $z_{j,j+1}=0.5\lambda$, $\Delta \omega_{j,j+1}=0.1\Gamma$, $\gamma=0$, and $\Gamma_{0}=1\Gamma$. The green dahsed-dotted line is the input photon spectrum, the red (blue) solid lines are the reflection (transmission) photon spectra. }
\end{figure*}

Our theory shown in Sec. II can be extended to calculate the single photon transport in a 1D waveguide coupled to arbitrary number of emitters. In this section, we take five emitters as an example. 

In the first example, we assume that the emitters are identical and the emitter separation is $0.05\lambda$. The emitter excitation dynamics and the emission spectra  are shown in Fig. 8(a) and 8(b), respectively. Here, we assume a single photon pulse with Gaussian shape is incident with $\Gamma_{0}=10\Gamma$ and when $\gamma=0.2\Gamma$. From Fig. 8(a), we see that the emitters can exchange excitations rapidly and the coherent population oscillations can last for an extended period of time. Similar to the two-emitter case, the coherent population oscillation is due to the coherent part of the dipole-dipole interactions between the emitters. From Fig. 8(b) we can see that there are five reflection peaks with two superradiant peaks on the higher frequency parts and three subradiant peaks on the lower frequency parts. This indicates that the collective interactions between the emitters split the single-excitation states into five eigenstates with two superradiant states and three subradiant states. This may be used as a frequency filter which can filter out some special frequencies.

In the second example, we consider the case that the emitters are not identical. These emitters have a spatial separation $0.5\lambda$ and the neighboring emitters have energy difference $\Delta \omega_{j,j+1}=0.1\Gamma$. The emitter excitation dynamics and the emission spectra when $\gamma=0$ are shown in Fig. 8(c) and 8(d) where we assume that the incident photon pulse has a Gaussian shape with $\Gamma_{0}=\Gamma$.  From Fig. 8(c), we see that the emitter excitations can oscillate and last for a very long time. The $j$th emitter and the $(N_{a}-j)$th emitter have almost the same excitation dynamics. The emission spectra are also very interesting. We can see that most of the photon spectra are reflected back but there are four very narrow transmission windows (Fig. 8(d)). This is the generalization of the DIET shown in previous sections. This phenomena may be used to generate a single photon frequency comb with very narrow linewidth \cite{Liao063004}.

\section{Summary}  

In summary, we have developed dynamical equations and photon emission spectra for a single photon transport in a 1D waveguide-QED system. In our generalized theory, the emitters can be either identical or non-identical. In addition, the dipole-dipole interactions induced by both the waveguide and non-waveguide vacuum modes are included. This theory allows one to calculate the real-time evolution of the photon pulse and the emitters in a 1D waveguide-QED system and study the many-body physics.

We first compare the results with and without including the dipole-dipole interaction induced by the non-waveguide photon modes. The emitter dynamics and the scattering spectrum can be significantly modified by the dipole-dipole interaction induced by the non-waveguide vacuum modes if the emitter separation is much smaller than the resonance wavelength. We introduce a quantity (spectrum difference) to study the effects of the dipole-dipole interaction induced by the non-waveguide vacuum modes.  We find that when the emitter separation is much smaller than the resonant wavelength ($|V_{12}^{(nw)}| > \Gamma $) the dipole-dipole interaction induced by the non-waveguide photon modes can considerably influence the photon dynamics. When the emitter separation is of the order of or larger than the resonant wavelength ($|V_{12}^{(nw)}| \ll  \Gamma $), the effects of the non-waveguide photon modes can be neglected. 

We then studied the case of non-identical emitters. The results show that if the energy difference between the emitters is much larger than the energy shift due to the dipole-dipole interaction ($\Delta \omega_{12}\gg 2\text{Im}(V_{12})$) the emitters behave like independent emitters. Otherwise, the emitters can strongly couple to each other. More interestingly, when the two emitters have close but non-zero energy difference, there is a  very narrow transparency window around the resonance frequency due to the interference between the two collective decay  channels. This is the demonstration of the dipole-dipole induced eltromagnetic transparency which may find important applications in quantum waveguide devices. For the case of multiple emitters, a single photon frequency comb with very narrow comb linewidth can be generated.

\section{Acknowledgment}
This work is supported by a grant from the Qatar National Research Fund (QNRF) under the NPRP project 7-210-1-032. 

\appendix
\numberwithin{equation}{section}

\section{Derivation of the emitter dynamical equations}

In this appendix, we derive the dynamical equations of the emitter system shown in Eq. (9). To derive Eq. (9), we need to calculate the second and third terms of Eq. (8).

For the second term of Eq. (8), the summation over $k$ can be replaced by an integration
\begin{equation}
\sum_{k}\rightarrow \frac{L}{2\pi}\int_{-\infty }^{\infty }dk, 
\end{equation}
where $L$ is the quantization length in the propagation direction. The second term of Eq. (8) can then be calculated as
\begin{widetext}
\begin{align}
&\sum_{k}g_{k}^{j}g_{k}^{l*}e^{ik(z_{j}-z_{l})}e^{i\delta\omega_{k}^{l}t'}e^{-i\delta\omega_{k}^{j}t} =\frac{L}{2\pi}\int_{-\infty }^{\infty }g_{k}^{j}g_{k}^{l*}e^{ik(z_{j}-z_{l})}e^{i\delta\omega_{k}^{l}t'}e^{-i\delta\omega_{k}^{j}t}dk   \\
&\simeq \frac{L}{2\pi}g_{k_a}^{j}g_{k_a}^{l*} e^{-i\Delta k_{l}v_{g}t'}e^{i\Delta k_{j}v_{g}t}\Big[\int_{0}^{\infty }e^{ik(z_{j}-z_{l})}e^{i(k-k_{a})v_{g}(t'-t)}dk+\int_{-\infty }^{0}e^{ik(z_{j}-z_{l})}e^{i(-k-k_{a})v_{g}(t'-t)}dk \Big ]    \\
&=\frac{L}{2\pi}g_{k_a}^{j}g_{k_a}^{l*} e^{-i\Delta k_{l}v_{g}t'}e^{i\Delta k_{j}v_{g}t}\Big\{e^{ik_{a}(z_j-z_{l})}\int_{-k_{a} }^{\infty }e^{i\delta k[(z_{j}-z_{l})+v_{g}(t'-t)]}d\delta k+e^{-ik_{a}(z_j-z_{l})}\int_{-k_{a} }^{\infty }e^{-i\delta k[(z_{j}-z_{l})-v_{g}(t'-t)]}d\delta k \Big\}   \nonumber \\ \\
&\simeq \frac{L}{2\pi}g_{k_a}^{j}g_{k_a}^{l*}e^{-i\Delta k_{l}v_{g}t'}e^{i\Delta k_{j}v_{g}t}\Big\{e^{ik_{a}(z_j-z_{l})}\int_{-\infty }^{\infty }e^{i\delta k[(z_{j}-z_{l})+v_{g}(t'-t)]}d\delta k+e^{-ik_{a}(z_j-z_{l})}\int_{-\infty }^{\infty }e^{-i\delta k[(z_{j}-z_{l})-v_{g}(t'-t)]}d\delta k \Big\} \nonumber \\ \\
&=Lg_{k_a}^{j}g_{k_a}^{l*} e^{-i\Delta k_{l}v_{g}t'}e^{i\Delta k_{j}v_{g}t}\Big\{e^{ik_{a}(z_j-z_{l})}\delta [(z_{j}-z_{l})+v_{g}(t'-t)]+e^{-ik_{a}(z_j-z_{l})}\delta [(z_{j}-z_{l})-v_{g}(t'-t)]\Big\}   \\
&=\frac{Lg_{k_a}^{j}g_{k_a}^{l*}}{v_{g}}e^{-i\Delta k_{l}v_{g}t'}e^{i\Delta k_{j}v_{g}t}\Big\{e^{ik_{a}(z_j-z_{l})}\delta [t'-(t-\frac{z_j-z_l}{v_{g}})]+e^{-ik_{a}(z_j-z_{l})}\delta [t'-(t+\frac{z_j-z_l}{v_{g}})]\Big\}   \\
&=\frac{\Gamma_{jl}}{2}e^{-i\Delta k_{l}v_{g}t'}e^{i\Delta k_{j}v_{g}t}e^{ik_{a}|z_j-z_l|}\delta [t'-(t-\frac{|z_j-z_l|}{v_{g}})] \\
&=\frac{\Gamma_{jl}}{2}e^{i\Delta \omega_{jl} t}e^{ik_{l}|z_{jl}|}\delta [t'-(t-\frac{|z_{jl}|}{v_{g}})]
\end{align}
\end{widetext}
where $\Delta \omega_{jl}=(k_{j}-k_{l})v_{g}$ is the energy difference between two emitters, $\Delta k_{j}=k_{j}-k_{a}$ with $k_{a}$ being a reference wavevector which can be chosen as the average wavevector (i.e., $k_{a}=\sum_{j}k_{j}/N_{a}$), and $\Gamma_{jl}=2Lg_{k_a}^{j}g_{k_a}^{l*}/v_{g}$.
From Eq. (A. 2) to Eq. (A. 3), we rewrite the integration into the left-propagation and right-propagation parts and assume that the coupling strength is uniform for the modes close to $k_{a}$. From Eq. (A. 4) to Eq. (A. 5), for $k_{a}\gg 0$ we can extend the lower bound of the integration from $-k_{a}$ to $-\infty$ and use the identity $\int_{-\infty}^{\infty}e^{ikx}dx=2\pi\delta(x)$. In Eq. (A. 7), since $t'\leq t$, when $r_{j}>r_{l}$ only the second term survives. On the contrary, when $r_{j}<r_{l}$ only the first term survives. Therefore, Eq. (A. 7) can be rewritten as Eq. (A. 8). By inserting Eq. (A. 9) into the second term of Eq. (8) we can obtain 
\begin{equation}
-\sum_{l=1}^{N}\frac{\Gamma_{jl}}{2}e^{i\Delta \omega_{jl} t}e^{ik_{l}|z_{jl}|}\alpha_{l}(t-\frac{|z_{jl}|}{v_{g}})
\end{equation} 
where $\Gamma_{i}=2L|g_{k_{a}}^{i}|^2/v_{g}$ is the decay rate of the $i$th emitter due to the guided photon modes and $|z_{jl}|=|z_{j}-z_{l}|$ is the emitter separation along the waveguide direction. 

To calculate the third term in Eq. (8), we first rewrite the summation over the wavevector $\vec{q}$ as an integration
\begin{equation}
\sum_{\vec{q}}=\frac{V}{(2\pi)^{3}}\int_{0}^{2\pi}d\phi\int_{0}^{\pi}\text{sin}\theta d\theta\int_{0}^{\infty}q^{2}dq,
\end{equation}
and the summation over the two polarizations as
\begin{equation}
\sum_{\lambda}g_{\vec{q},\lambda}^{j}g_{\vec{q},\lambda}^{l*}=\frac{\nu _{q}\mu _{ab}^j \mu _{ab}^l}{2\hbar\epsilon _{0}V}[(\hat{\mu}_{ab}\cdot \hat{e}_{\vec{q}}^{1} )^2+(\hat{\mu}_{ab}\cdot \hat{e}_{\vec{q}}^{2})^2],
\end{equation}
where $\nu _{q}$ is the photon frequency with wavevector $\vec{q}$,  $\mu _{ab}$ is the amplitude of the transition dipole moment with direction $\hat{\mu}_{ab}$, $\hat{e}_{\vec{q}}^{1}$ and  $\hat{e}_{\vec{q}}^{2}$ are the two polarization directions of the photon. Without loss of generality, we can assume the direction of the atomic transition dipole moment to be $\hat{\mu}_{ab}=(\text{sin}\varphi ,0,\text{cos}\varphi)$. The unit wavevector of the photon can be written as $\hat{q}=(\text{sin}\theta\text{cos}\phi,\text{sin}\theta\text{sin}\phi,\text{cos}\theta)$ and the two polarization directions are given by $\hat{e}_{\vec{q}}^{1}=(\text{sin}\phi,-\text{cos}\phi,0)$ and $\hat{e}_{\vec{q}}^{2}=(\text{cos}\theta\text{cos}\phi,\text{cos}\theta\text{sin}\phi,-\text{sin}\theta)$. Thus, we have
\begin{multline}
\sum_{\lambda}g_{\vec{q},\lambda}^{j}g_{\vec{q},\lambda}^{l*} \\ =\frac{\nu _{q}\mu _{ab}^{j}\mu _{ab}^{l}}{2\hbar\epsilon _{0}V}[\text{sin}^{2}\varphi \text{sin}^{2}\phi+(\text{sin}\varphi \text{cos}\theta\text{cos}\phi-\text{cos}\varphi\text{sin}\theta)^2].
\end{multline}

For $j=l$, using the Weisskoph-Wigner approximation and $\int_{-\infty}^{\infty}d\nu_{q}e^{i(\nu_{q}-\omega)(t'-t)}=2\pi\delta (t'-t)$, it is not difficult to obtain
\begin{equation}
\sum_{\vec{q},\lambda}|g_{\vec{q},\lambda}^{j}|^{2}\int_{0}^{t}\alpha_{j}(t')e^{i\delta\omega_{\vec{q}}^{j}t'}dt' e^{-i\delta \omega_{\vec{q}}^{j}t}=\frac{\gamma_j}{2}\delta_{jl}\alpha_{j}(t)
\end{equation}
where $\gamma_j=k_a^3\mu_{j}^{2}/3\pi\hbar\epsilon _{0}V$ is the spontaneous decay rate of the $j$th atom due to the non-waveguide photon modes. 

\begin{widetext}
For $j\neq l$, by integrating out the $\theta$ and $\phi$ we have
\begin{equation}
\sum_{\vec{q},\lambda}g_{\vec{q},\lambda}^{j}g_{\vec{q},\lambda}^{l*} e^{-i\vec{q}\cdot(\vec{r}_{j}-\vec{r}_{l})}=\frac{\nu_{k_{a}}\mu_{j}\mu_{l}}{4\pi^{2}\hbar \epsilon _{0}V}\int_{0}^{\infty} q^2\Big\{\text{sin}^2\varphi\frac{\text{sin}(q r_{jl})}{q r_{jl}}+(1-3\text{cos}^2\varphi)[\frac{\text{cos}(qr_{jl})}{(qr_{jl})^2}-\frac{\text{sin}(qr_{jl})}{(qr_{jl})^3}]\Big\}dq.
\end{equation}
where we assume that only a narrow band of frequency around resonant frequency can couple to the system, i.e., $\nu_{q}\simeq \nu_{k_{a}}$.
We then have
\begin{multline}
\sum_{\vec{q},\lambda}g_{\vec{q},\lambda}^{j}g_{\vec{q},\lambda}^{l*} e^{-i\vec{q}\cdot(\vec{r}_{j}-\vec{r}_{l})}e^{i\delta\omega_{q}^{jl}(t'-t)} \\
=\frac{\nu_{k_a}\mu_{j}\mu_{l}}{4\pi^{2}\hbar \epsilon _{0}V}e^{-i\Delta k_{l}v_{g}t'}e^{i\Delta k_{j}v_{g}t}\int_{0}^{\infty} q^2\Big\{\text{sin}^2\varphi\frac{\text{sin}(qr_{jl})}{qr_{jl}}+(1-3\text{cos}^2\varphi)[\frac{\text{cos}(qr_{jl})}{(qr_{jl})^2}-\frac{\text{sin}(qr_{jl})}{(qr_{jl})^3}]\Big\}e^{i(q-k_{a})v_{g}(t'-t)}dq
\end{multline}
The integration over the first term in the curly bracket can be calculated as follows
\begin{align}
&\int_{0}^{\infty}q^2\frac{\text{sin}(qr_{jl})}{qr_{jl}}e^{i(q-k_{a})v_{g}(t'-t)}dq
=\frac{1}{2ir_{jl}}\int_{0}^{\infty}q(e^{iqr_{jl}}-e^{-iqr_{jl}})e^{i(q-k_{a})v_{g}(t'-t)}dq \\
&\simeq \frac{k_{a}}{2ir_{jl}}\Big[\int_{-k_{a}}^{\infty}e^{ik_{a} r_{jl}}e^{i\delta q[r_{jl}+v_{g}(t'-t)]}d\delta q -\int_{-k_{a}}^{\infty} e^{-ik_{a} r_{jl}}e^{i\delta q[r_{jl}-v_{g}(t'-t)]}d\delta q \Big] \\
&\simeq \frac{k_{a}}{2ir_{jl}}\Big [\int_{-\infty}^{\infty}e^{ik_{a} r_{jl}}e^{i\delta q(r_{jl}+v_{g}(t'-t)}d\delta q -\int_{-\infty}^{\infty}e^{-ik_{a} r_{jl}}e^{i\delta q(r_{jl}-v_{g}(t'-t)}]d\delta q \Big ] \\
&=\frac{2\pi k_{a}}{2ir_{jl}v_{g}}\Big\{e^{ik_{a} r_{jl}}\delta[t'-(t-\frac{r_{jl}}{v_{g}})]-e^{-ik_{a} r_{jl}}\delta[t'-(t+\frac{r_{jl}}{v_{g}})]\Big\} \\
&=\frac{\pi k_{a}^{2}}{v_g}\cdot\frac{1}{ik_{a}r_{jl}}e^{ik_{a} |r_{jl}|}\delta[t'-(t-\frac{|r_{jl}|}{v_{g}})].
\end{align}
According to the Weisskopf-Wigner approximation \cite{Scully2001}, since the phase varies little around resonant frequency and it has the major contribution, from Eq. (A. 17) to Eq. (A. 18) we use $q\simeq k_{a}$ and move $q$ out of the integration, and change the lower bound of the integration from $-k_{a}$ to $-\infty$ from Eq. (A.18) to Eq. (A.19). Similarly, we have the second term and the third term of Eq. (A. 15) which are respectively given by
\begin{equation}
\int_{0}^{\infty} q^2\frac{\text{cos}(qr_{jl})}{(qr_{jl})^2}e^{i(q-k_{a})v_{g}(t'-t)}dq
=\frac{\pi k_{a}^{2}}{v_g}\cdot\frac{1}{(k_{a}r_{jl})^2}e^{ik_{a} |r_{jl}|}\delta[t'-(t-\frac{|r_{jl}|}{v_{g}})]
\end{equation}
and 
\begin{equation}
\int_{0}^{\infty} q^2\frac{\text{sin}(qr_{jl})}{(qr_{jl})^3}e^{i(q-k_{a})v_{g}(t'-t)}dq
=\frac{\pi k_{a}^{2}}{v_g}\cdot\frac{1}{i(k_{a}r_{jl})^3}e^{ik_{a} |r_{jl}|}\delta[t'-(t-\frac{|r_{jl}|}{v_{g}})]
\end{equation}
On inserting Eqs. (A. 21-A. 23)  into Eq. (A. 15), we have  
\begin{align}
&\sum_{\vec{q},\lambda}g_{\vec{q},\lambda}^{j}g_{\vec{q},\lambda}^{l*} e^{-i\vec{q}\cdot(\vec{r}_{j}-\vec{r}_{l})}e^{i\delta\omega_{q}^{l}t'}e^{-i\delta\omega_{q}^{j}t)} \\
&=\frac{3\gamma_{jl}}{4}e^{-i\Delta k_{l}v_{g}t'}e^{i\Delta k_{j}v_{g}t}\Big\{\text{sin}^{2}\varphi\frac{-i}{k_{a}r_{jl}}+(1-3\text{cos}^{2}\varphi)[\frac{1}{(k_a r_{jl})^2}+\frac{i}{(k_a r_{jl})^3}]\Big\}e^{ik_{a}|r_{jl}|}\delta[t'-(t-\frac{|r_{jl}|}{v_{g}})] \\
&=\frac{3\gamma_{jl}}{4}e^{-i\Delta \omega_{jl} t}e^{ik_{l}|r_{jl}|}\Big\{\text{sin}^{2}\varphi\frac{-i}{k_{a}r_{jl}}+(1-3\text{cos}^{2}\varphi)[\frac{1}{(k_a r_{jl})^2}+\frac{i}{(k_a r_{jl})^3}]\Big\}\delta[t'-(t-\frac{|r_{jl}|}{v_{g}})] 
\end{align}
On inserting Eqs. (A. 10) and (A. 26) into Eq. (8) we can obtain the dynamical equations of the emitters
\begin{equation}
\dot{\alpha}_{j}(t)=b_{j}(t)-\sum_{l=1}^{N} \Big[V_{jl}^{(w)}e^{ik_{l} |z_{jl}|} \alpha_{l}(t-\frac{|z_{jl}|}{v_{g}})+V_{jl}^{(nw)}e^{ik_{l} |r_{jl}|} \alpha_{l}(t-\frac{|r_{jl}|}{v_{g}})\Big ]e^{i\Delta \omega_{jl} t}, 
\end{equation}
where 
\begin{equation}
b_{j}(t)=-\frac{i}{2\pi}\sqrt{\frac{\Gamma v_{g}L}{2}}e^{ik_{a}z_{j}}e^{i\Delta k_{j}v_{g}t}\int_{-\infty }^{\infty }\beta_{\delta k}(0)e^{i\delta k(r_{j}-v_{g}t)}d\delta k,
\end{equation} 
$V_{jj}^{(w)}=\Gamma_{j}/2$ and $V_{jj}^{(nw)}=\gamma_{j}/2$. For $j\neq l $, $V_{jl}^{(w)}=\sqrt{\Gamma_{j}\Gamma_{l}}/2$ is the dipole-dipole coupling due to the waveguide modes and
\begin{equation}
V_{jl}^{(nw)}=\frac{3\sqrt{\gamma_{j}\gamma_{l}}}{4}\Big [\text{sin}^{2}\varphi\frac{-i}{k_{a}r_{jl}}+(1-3\text{cos}^{2}\varphi)\frac{1}{(k_{a}r_{jl})^2}+\frac{i}{(k_{a}r_{jl})^3} \Big ].
\end{equation}
\end{widetext}
is the dipole-dipole interaction due to the non-waveguide photon modes. The term $e^{-i\Delta \omega_{jl} t}$ is due to the energy difference between the two emitters. If the two emitters are the same, this term becomes unit and the equation returns back to the case for identical emitters.

\end{document}